%% file: main.tex
\documentclass[11pt,a4paper]{article}
\usepackage{jcappub}
\usepackage{color}
\usepackage{natbib}
\usepackage{subfigure}
\usepackage{verbatim}
\usepackage{graphicx}
\usepackage{bm}              
\usepackage[amssymb]{SIunits}
\usepackage{amssymb}
\usepackage{amsmath}
\usepackage{multirow}
\usepackage{tabularx}
\usepackage{booktabs}
\usepackage{float}
\usepackage[outercaption]{sidecap} 
\usepackage{threeparttable}

\usepackage{pifont}

\DeclareFixedFont{\ttb}{T1}{txtt}{bx}{n}{9} 
\DeclareFixedFont{\ttm}{T1}{txtt}{m}{n}{9}  

\newcommand{\ba}{\begin{eqnarray}}
\newcommand{\ea}{\end{eqnarray}}

\newcommand  \gtsim  {\lower.5ex\hbox{$\; \buildrel > \over \sim \;$}}
\newcommand  \ltsim  {\lower.5ex\hbox{$\; \buildrel < \over \sim \;$}}

\newcommand{\LCDM}   {$\Lambda$CDM}

\newcommand{\be}{\begin{equation}}
\newcommand{\ee}{\end{equation}}

\newcommand{\neff}  {$N_{\rm eff}$}
\newcommand{\mnu} {\Sigma m_\nu}

\bibpunct{(}{)}{;}{a}{}{,}

\def\3he{$^3{\rm He}$}

\def\PLANCK{{\sl Planck}}
\def\Euclid{{\sl Euclid}}
\def\Roman{{\sl Roman}}

\title{The Simons Observatory: Science Goals and Forecasts for the Enhanced Large Aperture Telescope}

\collaboration{The Simons Observatory Collaboration \note{Author contributions to this paper can be found at \url{https://simonsobservatory.org/wp-content/uploads/2025/03/Author-contribution-statement-20250228.pdf}.}}
\emailAdd{so\_publications@simonsobservatory.org}

\input{tex/author_JCAP_v4}

\abstract{We describe updated scientific goals for the wide-field, millimeter-wave survey that will be produced by the Simons Observatory (SO). Significant upgrades to the 6--meter SO Large Aperture Telescope (LAT) are expected to be complete by 2028, and will include a doubled mapping speed with 30,000 new detectors and an automated data reduction pipeline.  In addition, a new photovoltaic array will supply most of the observatory's power.  The LAT survey will cover about 60\% of the sky at a regular observing cadence, with five times the angular resolution and ten times the map depth of the {\it Planck} satellite. The science goals are to: (1) determine the physical conditions in the early universe and constrain the existence of new light particles; (2) measure the integrated distribution of mass, electron pressure, and electron momentum in the late-time universe, and, in combination with optical surveys, determine the neutrino mass and the effects of dark energy via tomographic measurements of the growth of structure at redshifts $z \lesssim 3$; (3) measure the distribution of electron density and pressure around galaxy groups and clusters, and calibrate the effects of energy input from galaxy formation on the surrounding environment; (4) produce a sample of more than 30,000 galaxy clusters, and more than 100,000 extragalactic millimeter sources, including regularly sampled AGN light-curves, to study these sources and their emission physics; (5) measure the polarized emission from magnetically aligned dust grains in our Galaxy, to study the properties of dust and the role of magnetic fields in star formation; (6) constrain asteroid regoliths, search for Trans-Neptunian Objects, and either detect or eliminate large portions of the phase space in the search for Planet 9; and (7) provide a powerful new window into the transient universe on time scales of minutes to years, concurrent with observations from the Vera C.~Rubin Observatory of overlapping sky.}

\toccontinuoustrue
\begin{document}
\maketitle

\section{Introduction and overall goals}
\label{sec:overview}
\input{tex/Introduction}

\section{Expanded capabilities}
\label{sec:infrastructure}
\input{tex/Infrastructure}

\section{Science goals and forecasts}
\label{sec:science}
\input{tex/Science_Drivers}

\section{Summary}
\label{sec:summary}
\input{tex/Summary}

\bibliography{SO-MSRI-2021,so}

\appendix
\section{Sensitivity and forecasting assumptions}
\label{app:sensitivities}
\input{tex/Sensitivities}

\end{document}

%% file: tex/author_JCAP_v4.tex
\author[1]{M.~Abitbol,}
\author[2,3]{I.~Abril-Cabezas,}
\author[4,5,6]{S.~Adachi,}
\author[7]{P.~Ade,}
\author[8,9]{A.~E.~Adler,}
\author[10]{P.~Agrawal,}
\author[11]{J.~Aguirre,}
\author[12,13]{Z.~Ahmed,}
\author[14]{S.~Aiola,}
\author[15,16]{T.~Alford,}
% ...

\affiliation[1]{Department of Physics, University of Oxford, Denys Wilkinson Building, Keble Road, Oxford OX1 3RH, United Kingdom} 
\affiliation[2]{DAMTP, Centre for Mathematical Sciences, University of Cambridge, Wilberforce Road, Cambridge CB3 OWA, UK} 
\affiliation[3]{Kavli Institute for Cosmology Cambridge, Madingley Road, Cambridge CB3 0HA, UK} 
\affiliation[4]{Hakubi Center for Advanced Research, Kyoto University, Kyoto 606-8501, Japan} 
\affiliation[5]{Department of Physics, Faculty of Science, Kyoto University, Kyoto 606-8502, Japan} 
\affiliation[6]{Kavli IPMU (WPI), UTIAS, The University of Tokyo, Kashiwa, Chiba 277-8583, Japan} 
\affiliation[7]{School of Physics and Astronomy, Cardiff University, UK} 
\affiliation[8]{Department of Physics, University of California Berkeley, Berkeley, CA, USA} 
\affiliation[9]{Lawrence Berkeley National Laboratory, Berkeley, CA, USA} 
\affiliation[10]{Department of Physics, University of California San Diego, San Diego, CA, USA} 
\affiliation[11]{Department of Physics and Astronomy, University of Pennsylvania, Philadelphia, PA 19104 USA} 
\affiliation[12]{Kavli Institute for Particle Astrophysics \& Cosmology, 452 Lomita Mall, Stanford, CA 94305, USA} 
\affiliation[13]{SLAC National Accelerator Laboratory, 2575 Sand Hill Road, Menlo Park, California 94025, USA} 
\affiliation[14]{Center for Computational Astrophysics, Flatiron Institute, USA} 
\affiliation[15]{University of Chicago, Department of Astronomy and Astrophysics, 5720 S Ellis Ave, Chicago, IL, 60637, USA} 
\affiliation[16]{University of Chicago, Department of Physics, 5720 S Ellis Ave, Chicago, IL, 60637, USA} 
% ...

%% file: tex/Introduction.tex
The millimeter-wave sky encodes information about the origins of the universe, the nature of gravity and other fundamental fields, the evolution of galaxies, the extent of our Solar System, and more. Unlocking this wealth of information requires large, well-characterized, multi-frequency maps with a large spatial dynamic range, high signal-to-noise, and good temporal coverage. The Simons Observatory (SO) is designed to make two surveys of the millimeter sky~\citep{so_forecast:2019}.  A deep, lower-resolution survey will search for primordial gravitational waves \citep{P52014,P52023}, either detecting them or producing an unprecedented limit on a measure of their amplitude, the tensor-to-scalar ratio. A wide, higher-resolution survey of 25,000 deg$^2$ reaching a co-added noise level\footnote{Throughout this work, K denotes CMB thermodynamic temperature units.} of 2.6~$\upmu$K$\cdot$arcmin will have ten times the co-added map depth of the \PLANCK\ satellite data with five times higher angular resolution~\citep{planck_overview:2018}, and will cover roughly sixteen times the area observed by the South Pole Telescope (SPT) 3G deep survey~\citep{2014SPIE.9153E..1PB,Prabhu2024}. Maps from the wide survey, made public on a regular schedule, can be used to address a broad set of questions in astronomy highlighted in the Decadal Survey on Astronomy and Astrophysics ``Astro2020'' White Papers, including the evolution of galaxies \citep{battaglia/etal:2019, dezotti/etal:2019b}, the role of magnetic fields in star formation in our Galaxy \citep{clark/etal:2019, fissel/etal:2019}, the properties of neutrinos and other light, relativistic particles \citep{alvarez/etal:2019, green/etal:2019, grin/etal:2019, grohs/etal:2019}, the characterization of primordial density perturbations \citep{gluscevic/etal:2019, meerburg/etal:2019, slosar/etal:2019b}, and beyond \citep{hensley/etal:2019, dezotti/etal:2019, holder/etal:2019, slosar/etal:2019}. 

The SO program is being enhanced in a number of ways beyond the capabilities described in~\cite{so_forecast:2019}. This paper focuses on the 6-meter Large Aperture Telescope (LAT) and the science enabled via its wide-area, high-resolution survey. The improved capabilities include doubling the number of detectors in the focal plane of the SO LAT receiver, developing a robust data pipeline that enables rapid mapmaking and transient alerts delivered to the community, and installing a photovoltaic array at the site on Cerro Toco, in the Atacama Desert of Chile, to provide power to the observatory. This expansion, and an extended duration of the SO survey through 2034, will significantly enhance the experimental capabilities and scientific return of the SO LAT. Other planned additions to the SO program, which target primordial gravitational-wave science with Small Aperture Telescopes \citep[SATs,][]{Galitzki2024}, will be described elsewhere. 

The remainder of this paper is organized as follows. In Sec.~\ref{sec:infrastructure}, we briefly describe the expanded capabilities of the wide-field LAT survey.   In Sec.~\ref{sec:science}, we forecast the primary scientific analyses that will be enabled by this program.  Much of the forecasting methodology used here is similar or identical to that employed in the SO science goals and forecasts paper~\citep{so_forecast:2019} or the complementary SO Galactic science goals and forecasts paper~\citep{SO_2022_Galactic_Science}, and we direct the interested reader to those papers for further details. 
In Sec.~\ref{sec:summary}, we conclude and discuss the outlook for SO operations and science.

\begin{table*}[t]
    \caption{SO Large Aperture Telescope Survey Specifications}
    \begin{tabular}{c | c | c | c | c | c | c }
    \hline
    \hline
        Frequency & FWHM & Baseline Depth & Goal Depth & Frequency & Detector & Optics \\
         $\mathrm{[GHz]}$ & $\mathrm{[arcmin]}$ & $[\upmu\mathrm{K} \cdot \mathrm{arcmin}]$ & $[\upmu\mathrm{K} \cdot \mathrm{arcmin}]$ & Bands & Count & Tubes \\
        \hline
        \begin{tabular}{@{}c@{}}
         27 (22 -- 30) \\ 39 (30 -- 47) \end{tabular} & \begin{tabular}{@{}c@{}} 7.4 \\ 5.1 \end{tabular} & \begin{tabular}{@{}c@{}} 61 \\ 30 \end{tabular} & \begin{tabular}{@{}c@{}} 44 \\ 23 \end{tabular} & LF & \begin{tabular}{@{}c@{}} 354 \\ 354 \end{tabular} & 1 \\
        \hline
        \begin{tabular}{@{}c@{}} 93 (77 -- 104) \\ 145 (128 -- 169) \end{tabular} & \begin{tabular}{@{}c@{}} 2.2 \\ 1.4 \end{tabular} & \begin{tabular}{@{}c@{}} 5.3 \\ 6.6 \end{tabular} & \begin{tabular}{@{}c@{}} 3.8 \\ 4.1 \end{tabular} & MF & \begin{tabular}{@{}c@{}} 20,640 \\ 20,640 \end{tabular} & 8 \\
        \hline
        \begin{tabular}{@{}c@{}} 225 (198 -- 256) \\ 280 (256 -- 313) \end{tabular} & \begin{tabular}{@{}c@{}} 1.0 \\ 0.9 \end{tabular} & \begin{tabular}{@{}c@{}} 15 \\ 35 \end{tabular} & \begin{tabular}{@{}c@{}} 10 \\ 25 \end{tabular} & UHF & \begin{tabular}{@{}c@{}} 10,320 \\ 10,320 \end{tabular} & 4 \\
        \hline
    \end{tabular}
\begin{tablenotes}
\item Expected instrumental and map-depth properties for the fully completed, nine-year SO LAT survey (2025-2034).  Two sensitivity targets are presented (baseline and goal), as in~\cite{so_forecast:2019}. The values in parentheses in the first column represent the approximate passband width of each frequency channel.
\end{tablenotes}
\label{tab:sens}
\end{table*}

%% file: tex/Infrastructure.tex
The mapping speed for the wide survey with the SO LAT~\citep{parshley/etal:2018,Gudmundsson2021} will be almost doubled by fully populating the LAT receiver \citep[LATR;][]{Zhu:2021} with six additional
optics tubes (OTs), thus adding roughly 30,000 detectors.\footnote{The initial seven OTs in the LATR comprise one low-frequency, four mid-frequency, and two ultra-high-frequency OTs~\citep{so_forecast:2019}.}  The six new OTs will comprise four mid-frequency (MF) OTs containing dichroic detectors with bands centered at 93~and 145~GHz \citep{Sierra:2025}, and two ultra-high-frequency (UHF) OTs containing dichroic detectors with bands centered at 225~and 280~GHz.  After installation of these new detectors, the receiver will contain eight MF, four UHF, and one low-frequency (LF) OTs, containing a total of roughly 60,000 detectors and fully populating the receiver's available focal plane.  The anticipated sensitivities of these detectors, including effects due to atmospheric noise, are described in \citet{so_forecast:2019}. The angular resolution of the LAT at these frequencies (e.g., FWHM $\approx$ 1.4~arcmin at 145~GHz) is also described there.  Table~\ref{tab:sens} summarizes the frequency channels, resolution, sensitivity, and detector counts for the fully populated LATR.  We note that these values reflect current best estimates derived from ongoing technical development of the SO LAT hardware, as well as the currently planned sky coverage of the SO LAT survey, which is wider than that assumed in~\cite{so_forecast:2019}.  Estimates of the bandwidth of each frequency channel are also provided, with each edge determined by the point at which the response drops to 50\% of the peak.  These are computed from the simulated optical coupling and on-chip filters.  Further discussion of the sensitivities can be found in Appendix~\ref{app:sensitivities}, including the per-OT noise-equivalent temperatures (NETs).

The lack of reliable power at the 5,200-meter site on Cerro Toco in Chile has been a risk to the performance of all projects operating there over the past few decades. The expanded SO program includes the installation of a photovoltaic array at the site, which will reduce the reliance on diesel-generated power by 70\% and provide a more stable power system.  This will translate into improved sensitivity by increasing on-sky observation time (we estimate a $5$--$10$\% increase in uptime), while also decreasing our environmental impact. 

The enhancement in hardware will be accompanied by substantial upgrades in the data processing pipeline to analyze and publicly deliver the data \citep[see][for recent work toward these goals]{Guan2024}. An open-source data pipeline will convert raw time-ordered data to maps of the sky and light curves of millimeter sources, to be released to the community.

\begin{figure*}[t]
\center \includegraphics[width=\linewidth]{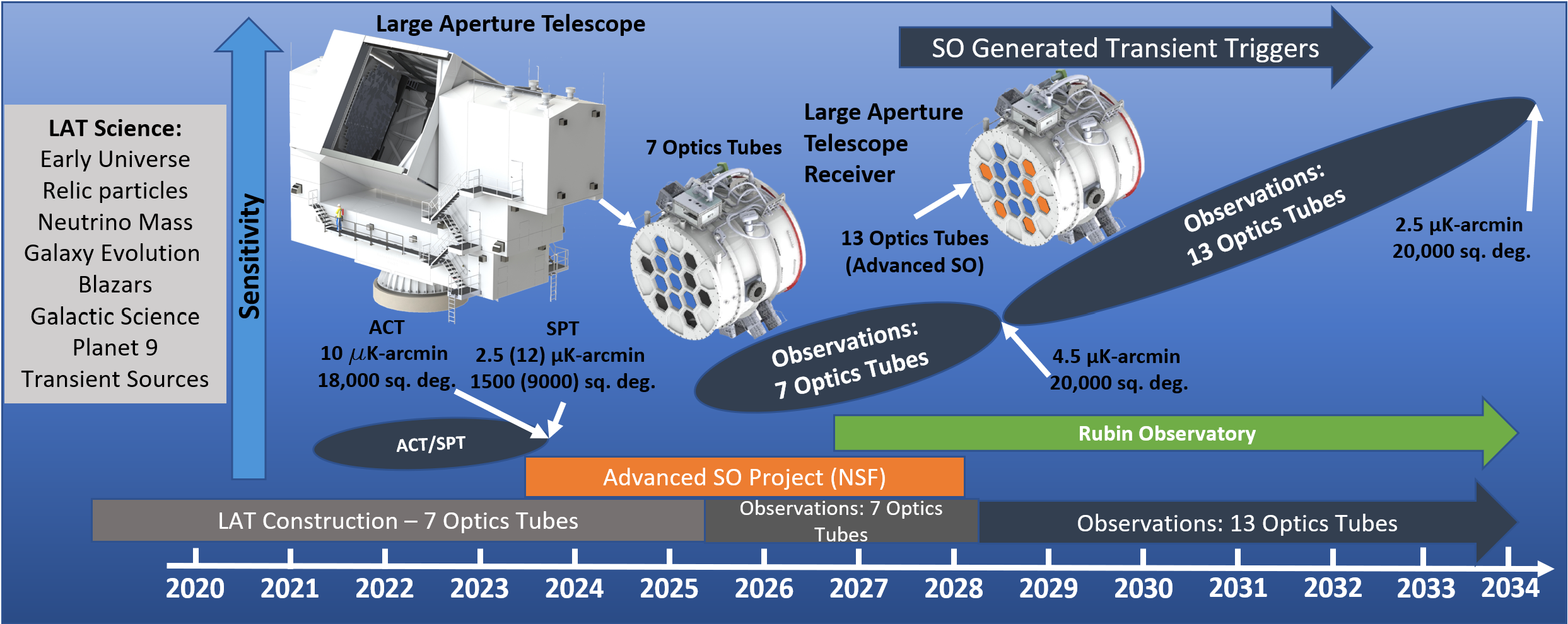}
\caption{Timeline of the Simons Observatory. 
The sensitivity, indicated by the overall vertical extent of each ellipse, approximates the survey-science reach of the different programs. 
The timing of SO and its broad sky coverage are ideally matched to the Rubin Observatory, DESI (2021-26), \Euclid\ (launched in July 2023), and {\sl Roman} (due to launch no later than 2027).  The 20,000 square degrees of overlapping sky coverage with Rubin is key to achieving the SO science goals.  A cutaway of the SO Large Aperture Telescope is shown, along with the Large Aperture Telescope Receiver.  The six new optics tubes described in Sec.~\ref{sec:infrastructure} are represented in orange on the front of the receiver.}
\label{fig:instrument} 
\end{figure*}

These new components are planned to be complete by 2028 (see timeline in Fig.~\ref{fig:instrument}), while the initial SO survey is ongoing using the half-populated LATR~\citep{Haridas2024}. Observations with the fully populated receiver will then commence, with a survey planned through 2034. The cumulative survey duration, including both the initial phase and the full-sensitivity phase, is expected to be nine years.  Note that the mapping speed of the LATR during the initial SO survey is roughly half that of the instrument during the latter part of the survey, after the upgrade. The forecasts presented in Sec.~\ref{sec:science} are based on the cumulative sensitivity of the entire survey with ``goal'' noise levels, accounting for the sensitivity improvement at the nominal-to-fully-populated LATR transition (see Appendix~\ref{app:sensitivities} for further details). The forecasts presented here include only statistical errors; systematic error contributions will be refined in future, dedicated studies.  In addition, \emph{Planck} data are incorporated in all cosmological forecasts unless explicitly stated otherwise --- \emph{Planck} is useful for measuring large-scale modes to which SO is less sensitive due to atmospheric noise, particularly in total intensity.  As a reference point, the depth of the co-added wide-area SO 93 and 145~GHz maps is expected to reach 2.8~$\upmu$K$\cdot$arcmin at the conclusion of the survey in 2034, assuming the ``goal'' noise levels in Table~\ref{tab:sens}.\footnote{Including all frequency maps, the depth is expected to reach 2.6~$\upmu$K$\cdot$arcmin; the LF and UHF channels are predominantly of use in foreground mitigation, rather than raw CMB sensitivity.}  Our forecasting approach follows that explained in detail in~\cite{so_forecast:2019}: we generate simulated sky maps of the microwave sky at all SO and \emph{Planck} frequencies, including all relevant sky components in temperature and polarization (with noise), and then process these through a harmonic-space internal linear combination pipeline to obtain post-component-separation noise power spectra, which are then used in all scientific forecasts.  The exceptions to this full-survey-integrated forecasting approach are the time-domain science cases.\footnote{The cosmic birefringence forecast in Sec.~\ref{sec:biref} also does not include {\sl Planck} data.} These include the analysis of AGN light curves in Sec.~\ref{sec:sources}, and millimeter-wave transient detection in Sec.~\ref{subsec:transients}, which both rely on the instantaneous sensitivity and transient detection pipeline of the SO survey with the fully populated LATR, rather than the cumulative depth of the full nine-year integrated data set.

%% file: tex/Science_Drivers.tex
\def\mnu{m_\nu}
\def\neff{$N_{\rm eff.}$}
\def\LCDM{$\Lambda$CDM}

\begin{figure*}[t]
\center 
\includegraphics[width=0.7\linewidth]{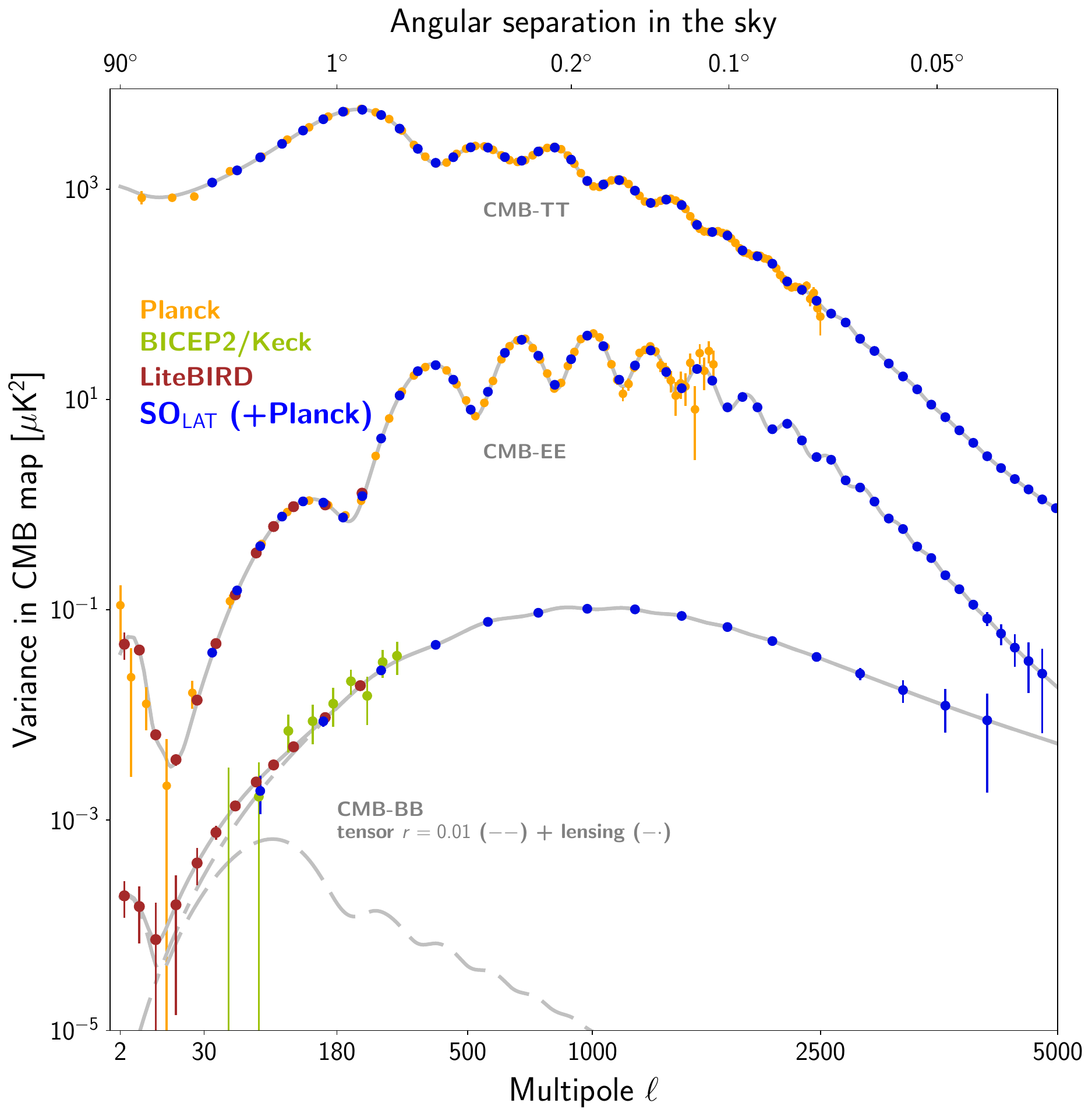}
\caption{Key CMB cosmology observables that can be derived from the SO LAT sky maps: CMB temperature (TT) and polarization (EE and BB) power spectra. The different colored points with error bars show existing (\PLANCK\ and BICEP2/Keck) and forecast ({\sl LiteBIRD} and SO) power spectrum measurements from various CMB experiments. The forecast SO noise power spectra here include detector and atmospheric noise, as well as the effects of residual foregrounds after component separation, following the methodology described in~\cite{so_forecast:2019}.  \PLANCK\ data are also assumed to be combined with SO, as indicated in the legend.  Small-scale power spectrum measurements from existing ground-based CMB experiments (e.g., ACT, SPT, POLARBEAR) are omitted here for clarity.}
\label{fig:science2} 
\end{figure*}

High-sensitivity, multi-frequency maps of the intensity and polarization of the  
millimeter-wave sky, observed at a regular cadence, will enable a broad set of new insights into our universe on scales ranging from the surface of last scattering to our Solar System.
The deployment timeline and key science forecasts are summarized in Fig.~\ref{fig:instrument} and Table \ref{tab:goals}, respectively. Fig.~\ref{fig:science2} gives an overview of the key CMB cosmology observables from the SO LAT survey. 
Much of the science requires the large sky coverage available from Chile (Fig.~\ref{fig:science_sky}).

\begin{table*}[t]
\caption[Simons Observatory Surveys]{Summary of Enhanced Science Goals from SO LAT Survey\textsuperscript{a}}  \footnotesize
\begin{tabular}{ l | c | c | c | l }
\hline
\hline

 & Current\textsuperscript{b} & {\bf SO} & Using Rubin, & Reference \\
 &                            & {\bf 2025--2034}             & DESI, or {\sl Euclid}    & \\
\hline
\hline
{\bf Primordial perturbations}  &&&&\\
\hline
$n_s$ & 0.004 & 0.002 &  - &\citet{shandera/etal:2019} \\
 $e^{-2\tau}\mathcal{P}(k=0.2\,\rm{Mpc^{-1}})$ & 3\% & 0.4\% &  - & \citet{slosar/etal:2019b}\\
$f^{\rm local}_{\rm{NL}}$ &  5&  1 & \ding{51} & \citet{meerburg/etal:2019}\\
\hline
{\bf Relativistic species} &&&&\\
\hline
$N_{\rm eff}$ &  0.2 & $0.045$  & - & \citet{green/etal:2019}\\
\hline
{\bf Neutrino mass\textsuperscript{c}} &&&&\\
\hline
$\sum \mnu$ (eV, $\sigma(\tau)=0.01$) & 0.1 &  0.03 & \ding{51}  & \citet{dvorkin/etal:2019} \\
$\sum \mnu$ (eV, $\sigma(\tau)=0.002$) & &   0.015  & \ding{51} &\\
\hline
{\bf Accelerated expansion} &&&&\\
\hline
$\sigma_8(z=1-2)$& 7\%& 1\% & \ding{51} & \citet{slosar/etal:2019}\\
\hline
{\bf Galaxy evolution} &&&&\\
\hline
$\eta_{\rm feedback}$ & 50--100\% & 2\% & \ding{51} &\citet{battaglia/etal:2019}\\
$p_{\rm nt}$  & 50--100\% & 4\% & \ding{51} & \citet{battaglia/etal:2019}\\
\hline
{\bf Reionization } &&&&\\
\hline
$\Delta z$ & 1.4 & 0.3 & - & \citet{alvarez/etal:2019}\\
$\tau$ & 0.007 & 0.0035 & - & \citet{alvarez/etal:2019}\\
\hline
{\bf Cluster catalog} & 4000 & 33,000 & \ding{51} &\\
{\bf AGN catalog} & 2000 & 96,000 & - &\\
\hline
{\bf Galactic science} &&&&\\
\hline
{Molecular cloud B-fields} & 10s & $>860$ & - & \citet{SO_2022_Galactic_Science}\\
{$\sigma (\beta_\mathrm{dust})$} & 0.02  & $0.005$ & - & \citet{SO_2022_Galactic_Science} \\
\hline
{\bf Solar System Science } &&&&\\
\hline
{Distance limit for 5 $M_\oplus$ Planet 9} & 500 AU &  900 AU  & \ding{51} & \citet{Fienga2020} \\
{Asteroid detections} &  &  $\sim10,000$  &  &\\
\hline
{\bf Transient detection } &&&&\\
{\bf \,\,\, distance} &&&&\\
\hline
{Long GRBs, on-axis} &  &  
1300 Mpc  & - &\\
{Low-luminosity GRBs} &  & 
70--210 Mpc  & - &\\
{TDEs, on-axis} &  & 670 Mpc  & - &\\
\hline
\end{tabular}
\begin{tablenotes}
\item \textsuperscript{a}  Projected 1$\sigma$ errors computed with standard methodology as in \citet{so_forecast:2019}, scaled to account for the improved noise from the enhanced infrastructure described in Sec.~\ref{sec:infrastructure} (see Appendix~\ref{app:sensitivities}).  Galactic science forecasts are computed as in \citet{SO_2022_Galactic_Science}. \citet{so_forecast:2019} describes our methods to account for noise properties and foreground uncertainties. We adopt ``goal'' noise levels for the SO LAT in these forecasts.  A 20\% end-to-end observation efficiency is used, matching that typically achieved in Chile \citep[as also assumed in][]{so_forecast:2019}. We assume the {\it Planck} data are included throughout. External data listed in the fourth column are those necessary to achieve the forecasted precision on each individual science target; for the cluster catalog the external data are needed only for obtaining redshifts.  Note that the time-domain science forecasts are new to this work, as this topic was not considered in the nominal SO science goals forecasting~\citep{so_forecast:2019}. 
\item \textsuperscript{b} Primarily from \PLANCK\ \citep{planck_cosmo:2018}. We anticipate constraints from existing ground-based data to improve on the ``current'' limits in the near future.  These constraints are expected to lie between the ``current'' and SO levels.
\item \textsuperscript{c} The forecast precision on $\sum m_\nu$ is highly sensitive to the projected error bar on $\tau$, and thus we provide two forecasts here (see Sec.~\ref{subsec:LSSmapping}).
\end{tablenotes}
\label{tab:goals}
\end{table*}

\subsection{Constraining the properties of primordial perturbations} 
\label{subsec:primordial}

By measuring primordial fluctuations in the CMB over twice the range of angular scales probed by the \PLANCK\ satellite, the full SO survey (in combination with \PLANCK) will significantly improve characterization of the scale dependence, Gaussianity, and adiabatic nature of the primordial density fluctuations that are the signature of the dynamics of the first moments of the universe \citep{meerburg/etal:2019,slosar/etal:2019b,hanany/etal:2019}. SO will halve the current error bar on the scalar perturbation spectral index $n_s$, testing the near-scale-invariant prediction of inflation over a wider range of scales than accessible to {\it Planck} \citep{slosar/etal:2019b}. SO will further test early-universe models by constraining the Gaussianity of the perturbations to $\sigma(f^{\rm local}_{\rm NL}) = 1$ via kinematic Sunyaev-Zel'dovich (kSZ) tomography \citep{Smith2018,Munchmeyer2019}, improving current constraints by a factor of five \citep{meerburg/etal:2019}, and also by constraining primordial isocurvature perturbations.\footnote{Direct estimates of $f^{\rm local}_{\rm NL}$ from the primary CMB bispectrum will also provide robust bounds, albeit with error bars roughly 2-3 times larger than those expected from kSZ tomography.  The primary bispectra will also tightly constrain other shapes of non-Gaussianity (equilateral and orthogonal), with roughly a factor of two improvement over the current bounds from \PLANCK.}  Note that achieving $\sigma(f^{\rm local}_{\rm NL}) = 1$ via kSZ tomography requires overlapping galaxy survey data, as will be available on the SO footprint from Rubin, {\sl Euclid}, and other surveys.  Constraints on primordial tensor perturbations, which require data from the SATs, will be described in a separate publication, which will also describe delensing forecasts enabled by the enhanced LAT infrastructure (see \cite{2022PhRvD.105b3511N} and \cite{2024PhRvD.110d3532H} for delensing forecasts for the nominal SO survey).  Together, these measurements will provide the most detailed constraints to date on the primordial power spectrum and early-universe physics.

\subsection{A refined image of the earliest snapshot of the universe}
\label{subsec:dampingtail}

The SO survey of the millimeter-wave sky will provide numerous new insights into fundamental physics.  Wide classes of beyond-standard-model (BSM) particle physics scenarios \citep[e.g., some scenarios constructed to solve known theoretical issues like the hierarchy problem;][]{2016PhRvL.117y1801A}, predict the existence of new light ($\lesssim 1$ eV) species that were in thermal equilibrium at some early time with the primordial plasma \citep{Essig2013,Alexander2016,green/etal:2019,snowrelic}.  Such scenarios leave distinct imprints in the small-scale damping tail of the CMB temperature and polarization anisotropy power spectra, thus yielding tight CMB-derived constraints on many BSM scenarios, including axions, sterile neutrinos, gravitinos, high-frequency gravitational waves, and other forms of relativistic energy density in the early universe. SO will either find evidence for new particles via these signatures or constrain BSM theories by improving current limits on the number of relativistic species \citep{planck_cosmo:2018} by a factor of four, with $\sigma(N_{\rm eff}) = 0.045$ (see Fig.~\ref{fig:science2} and Table~\ref{tab:goals}).  For example, the full SO survey will rule out at $>95$\% CL any light spin-3/2 particle that was in thermal equilibrium at any time back to reheating, assuming a standard thermal history of the universe after the particle's decoupling.  In particular, SO will significantly improve the precision with which the characteristic phase shift imprinted by free-streaming particles can be detected in the CMB, a unique signature sensitive to BSM physics~\citep{Montefalcone:2025unv}. 
A robust detection of $N_{\rm eff}$ differing from its standard-model value (3.044) would be landmark evidence for new physics, yielding the first direct cosmic signal from the epoch between post-inflationary reheating and neutrino decoupling one second later.  Importantly, $N_{\rm eff}$ is a generic, model-independent probe of new light particles in the early universe, thus yielding robust constraints on new physics across a vast search space~\citep{green/etal:2019,snowrelic}.

The reported tension between local and cosmological measurements of the rate of expansion of the universe \citep[as characterized by the Hubble constant, $H_0$;][]{verde/etal:2019,divalentino/etal:2020b,2024ApJ...973...30B,2024arXiv240806153F} may be resolved by new particle physics models that can increase the Hubble constant while preserving the fit to current CMB power spectrum data \citep[see, e.g.,][for a review]{2021CQGra..38o3001D}.  Models that alter the pre-recombination dynamics are of particular theoretical interest~\citep{KnoxMillea2020}, and generically imprint signatures in the CMB temperature and polarization on small scales.  Models that accelerate the recombination process itself are also of interest in this context~\citep[e.g.,][]{Chiang2018,H0Olympics}.\footnote{We note that detailed knowledge of the recombination process is crucial for the interpretation of CMB data~\citep{Planck2015params}.  In this context, the refined recombination codes CosmoRec~\citep{2011MNRAS.412..748C} and HyRec~\citep{2011PhRvD..83d3513A} are designed to reach sufficient accuracy for modeling standard recombination in the analysis of SO data.}  The full SO survey data will discriminate among such models \citep[e.g.,][]{Smith2020,Hill2022,Galli2022,Lynch2024,Kou2025}, or potentially make a detection.  The SO data will also enable sensitive searches for interacting dark matter particles, ultra-light axions, cosmic strings, and primordial magnetic fields, as well as precision tests of Big Bang Nucleosynthesis~\citep{gluscevic/etal:2019,grin/etal:2019,grohs/etal:2019}.

\subsection{Improving constraints on cosmic birefringence}
\label{sec:biref}

The SO data can be used to search for parity-breaking BSM physics, which can produce cosmic birefringence in the linear polarization of CMB photons~\citep{Carroll1998,Lue1999}. A canonical model generating such an effect is that of a new pseudo-scalar field coupled to the electromagnetic field-strength tensor via a Chern-Simons term~\citep{Sikivie1983,TurnerWidrow1988}.  A potential detection of such a field would have profound implications~\citep[e.g.,][]{Marsh2016,Ferreira2021}. This cosmic birefringence manifests as a rotation of the plane of linear polarization, giving rise to a nonzero $EB$ power spectrum in the CMB.  Strong limits have been placed on the $EB$ polarization angle with ACT data \citep[$0.07 \pm 0.09^\circ$;][]{choi2020}. Exploiting Galactic foregrounds to break the degeneracy between a miscalibrated instrumental polarization angle and actual cosmological rotation, recent analyses have reported hints of isotropic cosmic birefringence in \PLANCK\ data~\citep{MinamiKomatsu2020}, with the latest results estimating a birefringence angle $\beta = 0.342^{\circ} \substack{+0.094^\circ\\-0.091^\circ}$~\citep{Eskilt2022}.

The SO LAT survey will observe a significant fraction of the Galactic plane, which will allow us to test this methodology and derive an independent, significantly tighter constraint~\citep{Diego-Palazuelos2022,Diego-Palazuelos2023}. Using two detector splits for each of the six LAT frequency channels over the full survey region, and in the multipole range $\ell=100-2000$, the full SO survey will yield $\sigma(\beta)=0.04^{\circ}$, more than a factor of two improvement over current error bars, if the methodology and assumptions of \citet{Eskilt2022} are applied to these data. Note that this forecast uses only SO LAT data, i.e., no {\sl Planck} data are included, in contrast to the other cosmological forecasts in this paper.\footnote{Note that $1/f$ noise is included in the SO LAT noise model here, as in all other forecasts.}
Polarized thermal dust in the Milky Way shows a positive parity-violating $TB$ correlation~\citep{Planck2018:XI, Weiland:2020}, limiting current birefringence constraints due to the need to account for this poorly understood, parity-violating foreground; our forecast here assumes this ``intrinsic'' foreground contribution to the observed $EB$ correlation to be negligible, but this assumption may not hold~\citep{Clark_2021}. Efforts are ongoing to understand the mechanisms by which dust could produce this parity-violating signal~\citep[e.g.,][]{Huffenberger_2020,Clark_2021,Cukierman_2023,Diego-Palazuelos2023,Vacher_2023,Halal:2024, Hervias_2024}. SO and complementary surveys \citep[e.g., with the Fred Young Submillimeter Telescope;][]{CCAT2023} will enable characterization of the dust signal at high resolution and sensitivity, allowing improved modeling and more robust constraints on birefringence. Should a robust $EB$ signal be detected with SO, it will be important to account for lensing-induced smoothing of the $EB$ power spectrum in order to derive unbiased constraints on BSM physics~\citep{Naokawa2023,IdicherianLonappan:2025trj}.

\begin{figure*}[t]
\center \includegraphics[width=\linewidth]{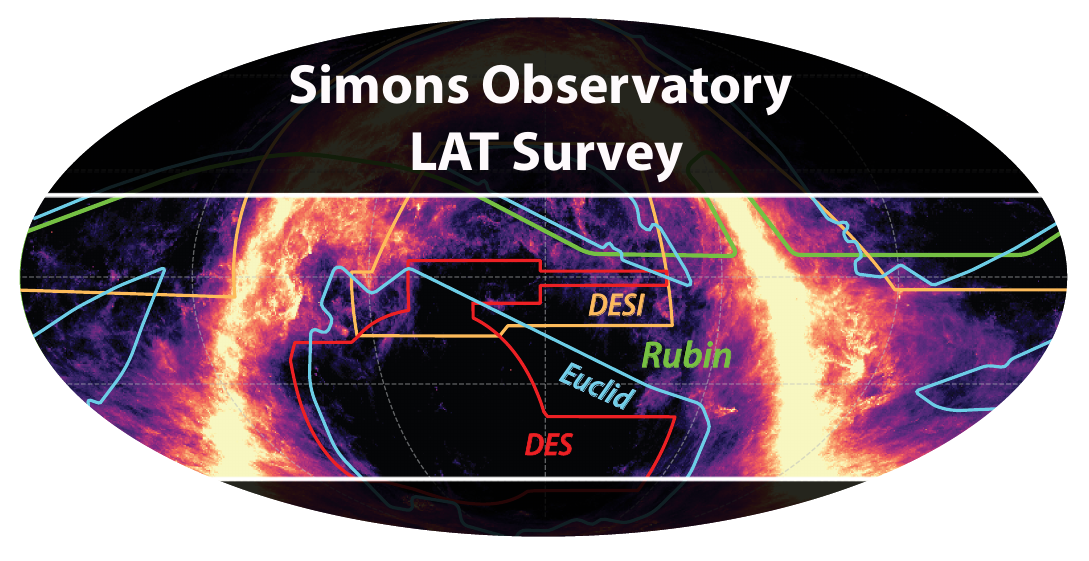}
\caption{The SO LAT survey (highlighted region with white boundaries) will cover 61\% of the sky. This sky coverage overlaps that of the Rubin Observatory and much of DES, DESI, and \Euclid, as shown. The background image shows the \textit{Planck} 857\,GHz total intensity map on a logarithmic scale.}
\label{fig:science_sky}
\end{figure*}

\subsection{A new large-scale view of dark matter, baryons, and galaxy clusters} 
\label{subsec:LSSmapping}

CMB photons interact with matter in the universe as it evolves over cosmic time.  Detailed analyses of the deflection of photons due to gravitational lensing from large-scale structures and the scattering of photons from ionized gas in galaxies, groups, and clusters offer exquisite tests of structure formation models and allow constraints on the physical parameters governing these processes, particularly via cross-correlation measurements, as enabled by the SO LAT survey sky coverage shown in Fig.~\ref{fig:science_sky}.  Here we describe a series of science goals that are enabled by measurements of these effects.

Neutrino oscillation experiments show the three neutrino species have a total mass of at least $0.06$~eV, or at least $0.1$~eV if the hierarchy of particle masses is inverted \citep{Agashe:2014kda}. In combination with DESI baryon acoustic oscillation data~\citep{2024arXiv240403002D}, the maps enabled by the enhanced infrastructure described in Sec.~\ref{sec:infrastructure} will provide a measurement with $\sigma(\sum\mnu)=0.03$~eV~\citep{so_forecast:2019}.  Combined with future improved constraints on the reionization optical depth, $\tau$~\citep[e.g., from {\sl LiteBIRD,}][]{litebird:2019}, the SO + DESI measurement is expected to improve to $\sigma(\sum\mnu)=0.015$~eV.\footnote{We also anticipate competitive $\tau$ constraints from SO itself via the patchy kinematic Sunyaev-Zel'dovich effect, as explained later in this section; although these will not reach the precision expected from {\sl LiteBIRD}, they will also improve the SO-derived neutrino mass constraint over that obtained with \PLANCK\ constraints on $\tau$.  Anticipated improvements on $\tau$ from CLASS~\citep{CLASS:2016} will also help in this regard.} Evidence for non-zero neutrino mass from cosmological data, and the resulting constraints on the hierarchy, will be of significant consequence to particle physics, as detailed in the 2014 and 2023 P5 reports~\citep{P52014,P52023}. 

The nature of dark energy is one of the most profound questions in modern physics and is a key science target for many ongoing and upcoming wide-area optical surveys, including the Dark Energy Survey~\citep{2016MNRAS.460.1270D}, the Hyper Suprime-Cam Survey~\citep{2022PASJ...74..247A}, the Rubin Legacy Survey of Space and Time~\citep{lsstbook}, \Euclid~\citep{Euclid_ScienceBook, EuclidCollaboration:2024}, \Roman~\citep{2015arXiv150303757S}, and others. Precision gravitational lensing measurements of the CMB by the fully populated SO LATR will complement optical observations from Rubin and \Euclid\ by extending the redshift range over which we can measure the effects of dark energy to $z>1$~\citep{Fang2021}.  This will probe earlier epochs of cosmic history than accessible to optical telescopes, which are limited by the faintness of distant galaxies, and will enable novel tests of modified gravity theories~\citep{slosar/etal:2019}.  Concretely, we forecast a 1\% constraint on the amplitude of the matter power spectrum at $z=1-2$ via the combination of SO CMB lensing maps with galaxy catalogs from Rubin and \Euclid, through combined analyses of CMB lensing auto- and cross-power spectra with these surveys.  These constraints will precisely probe dark energy at the epoch when its dynamical influence first becomes evident.  In addition, tomographic measurements of the growth of structure over a wide redshift range \citep[out to $z \approx 3$ and perhaps beyond;][]{Schmittfull2018,Qu2023} will also be enabled by similar cross-correlations with these and other surveys, including \Roman\ and {\sl SPHEREx}~\citep{SPHEREx}.

\begin{figure*}[t]
\center \includegraphics[width=\textwidth]{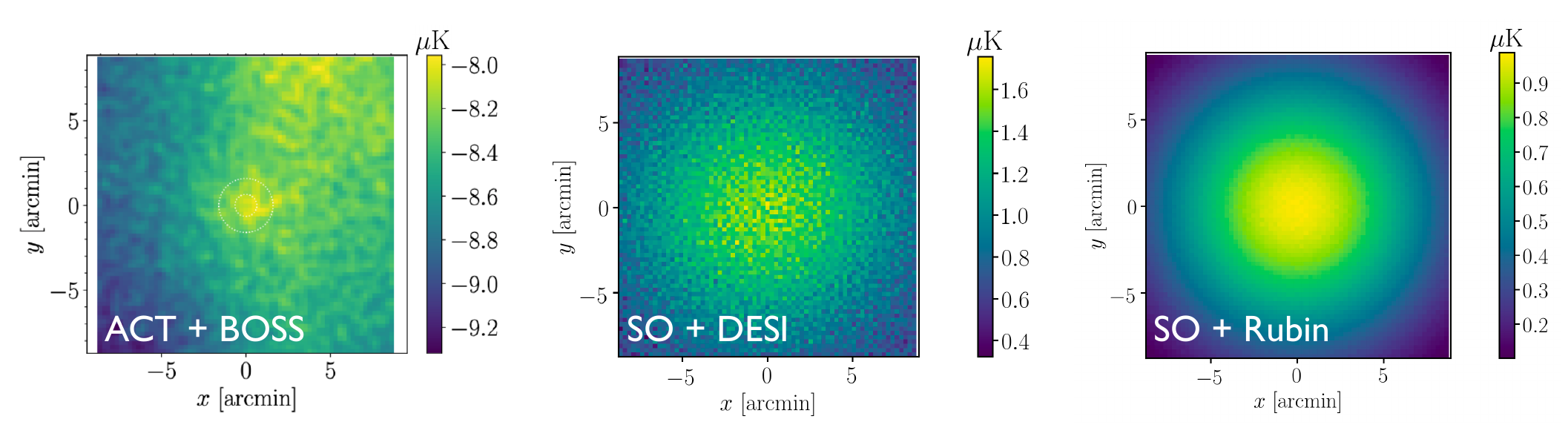}
\vspace{-0.2in}
\caption{Stacked images of ionized gas density around galaxies measured via the kSZ effect using ACT and BOSS data~\citep[left;][]{schaan/etal:2020}, simulated full-SO-survey and DESI data (middle), and simulated full-SO-survey and Rubin data (right).  Note that the left image has an overall additive offset and gradient due to residual CMB fluctuations in the stack.  The enhanced SO infrastructure will improve the precision of such measurements by nearly two orders of magnitude over current data.}
\label{fig:kSZtSZ} 
\end{figure*}

Characterizing the energy efficiency of feedback in galaxies, sourced by supernovae and active galactic nuclei, is a major focus of galaxy evolution studies. Feedback is also the dominant physical uncertainty in optical gravitational lensing analyses that seek to constrain dark energy. Combined with measurements of large-scale structure from DESI, Rubin, and \Euclid, the SO maps of integrated electron pressure (Compton-$y$) and momentum from the thermal and kinematic Sunyaev-Zel'dovich (SZ) effects~\citep{SZ1969,SZ1972,SZ1980} will uniquely measure the baryon content in galaxies, groups, and clusters -- a quantity not measurable with optical telescopes -- and will constrain the feedback efficiency to the few percent level.  
Fig.~\ref{fig:kSZtSZ} shows simulated stacked images of the ionized gas density in the circumgalactic medium around DESI and Rubin galaxies,\footnote{For DESI, we assume 2.9 million luminous red galaxies in this stack, measured on an overlapping sky area of 10,000~${\rm deg}^2$; for Rubin, we assume 1.5 billion galaxies measured on 16,000~${\rm deg}^2$.} as measured using the kSZ effect with the full SO survey data.  These measurements will improve over the precision of recent data from ACT and BOSS \citep[][also shown in Fig.~\ref{fig:kSZtSZ}]{schaan/etal:2020} by nearly two orders of magnitude \citep{Battaglia2017} (see also, e.g.,~\cite{2024arXiv240707152H} for recent kSZ measurements using ACT and DESI and \cite{Coulton2024ymap,Bleem2022,McCarthyHill2024,Tanimura2022,Chandran2023,2016A&A...594A..22P} for state-of-the-art Compton-$y$ maps from ACT, SPT, and {\sl Planck} data).

The SO maps will complement 21-centimeter experiments that probe the epoch of reionization \citep[e.g.,][]{2023ApJ...945..124H,2013PASA...30....7T,2013ExA....36..235M} by distinguishing among different models for how the universe was heated by the first ionizing sources. The ``patchy'' kSZ signal imprinted in the CMB during this epoch, due to scattering off the newly freed electrons, will enable the duration of the epoch of reionization to be measured with an uncertainty of $\sigma(\Delta z) = 0.3$ \citep[about 60 million years;][]{alvarez/etal:2019}.  This signal, interpreted using astrophysical reionization modeling, will also allow a determination of the optical depth $\tau$, independent of large-scale CMB $E$-mode measurements \citep{alvarez/etal:2020}.  We forecast that SO can reach $\sigma(\tau) = 0.0035$ using this new method, which relies on a joint analysis of the kSZ 2- and 4-point functions~\citep{alvarez/etal:2020,ferraro/smith:2018}.\footnote{For the purposes of this forecast, foregrounds are treated as Gaussian; future work will refine this methodology~\citep{Maccrann2024,Raghunathan2024}.}  This forecast is within a factor of two of the cosmic-variance-limited error bar from the large-scale $E$-mode polarization targeted by satellite missions such as {\sl LiteBIRD} \citep{LiteBIRDCollaboration:2023}.

Using the unique redshift independence of the thermal SZ (tSZ) effect, SO will extend tSZ cluster detection into the epoch in which the first massive, virialized structures formed at $z \gtrsim 2$.  Thermal SZ surveys from ACT, SPT, and \textit{Planck} have led the field in constructing clean, complete, nearly mass-selected cluster samples, with the latest catalogs comprising several thousand galaxy clusters in total \citep{Hilton2021,SPT:2023tib,Planck:2015koh}. We forecast that the full SO survey will detect $33,000$ clusters, with redshifts from overlapping DESI, Rubin, \Euclid, and {\sl SPHEREx} data, as well as dedicated optical and near-IR observations.  This forecast uses the multifrequency matched-filter methodology described in~\cite{Madhavacheril2017} \citep[following][]{Herranz2002,Melin2006} assuming the gas pressure profile from~\cite{Arnaud2010}\footnote{The exact pressure profile parameters used are: $P_0 = 8.403$, $c = 1.156$, $\alpha = 1.062$, $\gamma = 0.3292$, and $\beta = 5.4807$, with a hydrostatic mass bias $1-b = 0.8$ used to set the normalization of the pressure--mass relation.} with a cut on SNR $> 5$ used to define the cluster sample.  SO will thus provide the 
community with a homogeneous, well-defined catalog for follow-up studies out to high redshifts, with $\approx200$ clusters in the unique $z > 2$ discovery space~\citep{mantz/etal:2019}, enabling a broad range of multiwavelength science, including X-ray analyses with {\sl eROSITA}~\citep{2024A&A...682A..34M}.  Mass estimates will be enabled by galaxy weak lensing data from Rubin and \Euclid\ for clusters at $z \lesssim 1$ and by SO's own CMB lensing data at higher redshifts.  SO measurements of high-$z$ clusters and proto-clusters will also provide crucial total flux constraints for high-resolution follow-up observations with interferometric facilities (e.g.,~\citet{2023PASJ...75..311K,2024A&A...689A..41V}). 

\begin{figure}[t]
\center
\includegraphics[width=0.65\textwidth]{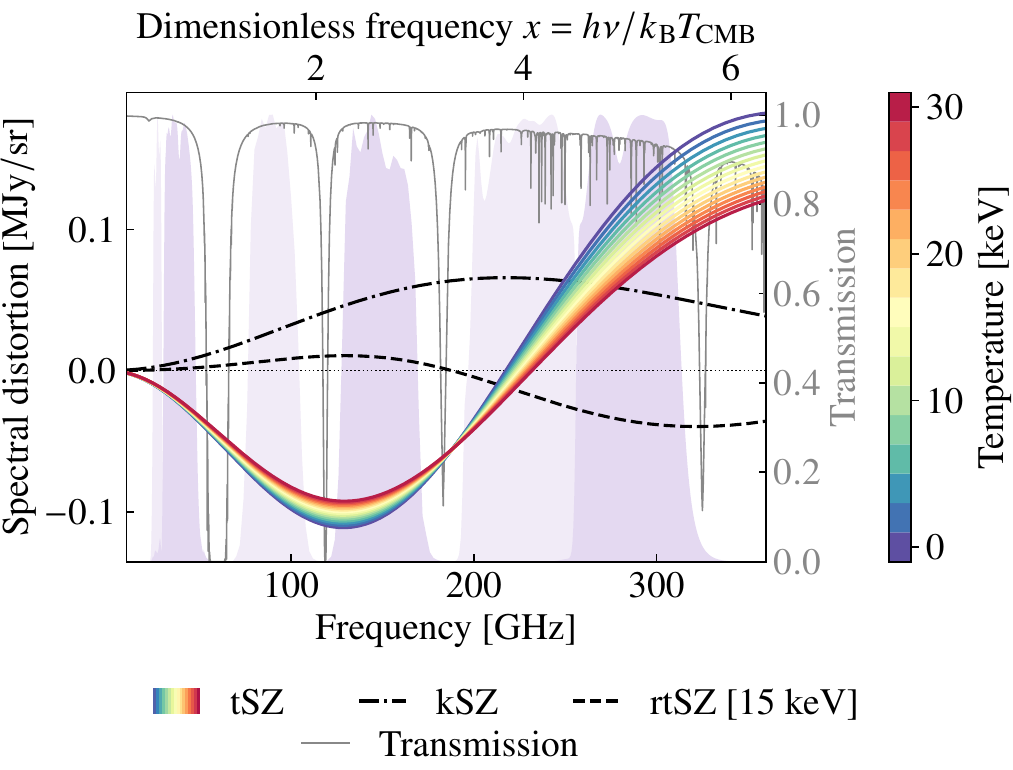}
\caption{Spectral dependence of the intensity of the kinematic SZ (dot-dashed curve) and relativistically corrected thermal SZ effects, where the line color denotes the electron temperature as labeled.  The relativistic corrections are computed using {\tt SZpack}~\citep{Chluba2012SZpack,Lee2024}, with the difference between the thermal SZ spectrum for a 15~keV plasma versus that for the non-relativistic case shown as a dashed curve.  The background shaded purple regions show the six bands of the SO LATR~\citep{Simon:2018, Walker:2020, Sierra:2025}, and the light gray curve, computed using the {\tt am} code~\citep{Paine2019}, shows the atmospheric transmission for median conditions at the SO site.}
\label{fig:SZspec} 
\end{figure}

In addition, the six frequency channels of the SO LATR will enable measurement of the relativistic tSZ (rtSZ) effect, particularly for massive clusters, building upon the recent $3.5\sigma$ detection from ACT~\citep{Coulton2024rtSZ} \citep[see also earlier $\approx 2\sigma$ hints using {\sl Planck} data in][]{2018MNRAS.476.3360E,2024arXiv241002488R}.  Example rtSZ spectra computed with {\tt SZpack}~\citep{Chluba2012SZpack,Lee2024} are shown in Fig.~\ref{fig:SZspec}, along with the kSZ spectrum, the SO LATR bands~\citep{Zhu:2021}, and the atmospheric transmission at the SO site.  The spectra in the plot are computed for an example cluster with Compton-$y$ parameter $y = 10^{-4}$, optical depth $\tau = 0.01$, and line-of-sight peculiar velocity $v_{\rm LOS}/c = 0.005$.  Relativistic tSZ measurements will yield simultaneous inference of the electron pressure and temperature (and hence density) in massive halos, thereby allowing new approaches to cluster mass estimation and novel constraints on intracluster medium physics, particularly in combination with X-ray data \citep[see, e.g.,][for a review]{2019SSRv..215...17M}.

\subsection{A wealth of extragalactic sources: time-variable blazars and dusty galaxies}
\label{sec:sources}

Recent measurements of extragalactic source counts from ACT and SPT \citep{Everett:2020, Gralla:2020, Vargas:2023} are broadly consistent with AGN and dusty star-forming galaxy (DSFG) models, including the \cite{Lagache:2020} AGN and \cite{Cai:2013} DSFG models, at SO frequencies between 93 and 280 GHz. These models, combined with the noise levels in Table~\ref{tab:sourcesens}, indicate that 93 GHz will be the most sensitive frequency for detecting AGN, with an expected count of approximately 96,000 at the goal sensitivity (see Appendix \ref{app:sensitivities} for more details on the sensitivity estimation). The $5\sigma$ source sensitivity with a matched-filter source finder is 2.0 mJy over a sky fraction of 0.52. This sky fraction corresponds to the SO LAT survey region, excluding the Galactic plane based on the 80\% sky fraction \textit{Planck} Galactic emission mask. 

\begin{table*}[t]
    \centering  
    \begin{tabular}{c|c |c|c | c}
    \hline\hline
      Frequency & \multicolumn{2}{c|}{Single Observation Sensitivity} & \multicolumn{2}{c}{Co-added Sensitivity} \\
      
      [GHz] & \multicolumn{2}{c|}{[mJy]} & \multicolumn{2}{c}{[mJy]}\\
      \hline

       & Baseline & Goal & Baseline & Goal \\

       \hline

      27 & 37 & 27 & 2.8 & 2.3 \\

      39 & 25 & 19 & 1.8 & 1.5 \\

      93 & 9.5 & 6.9 & 0.52 & 0.40 \\

      145 & 13 & 8.3 & 0.67 & 0.50 \\

      225 & 26 & 17 & 1.4 & 0.98 \\

      280 & 49 & 34 & 2.5 & 1.8 \\
      
    \hline
    \end{tabular}
     \caption{Sensitivity of the fully populated LATR to point sources, specified as $1\sigma$ root-mean-square (RMS) noise. The single observation sensitivities are relevant for transient detections and the co-added sensitivities correspond to those obtained by the end of the survey. For each we report baseline and goal sensitivities.   
     }
    \label{tab:sourcesens}
\end{table*}

The most sensitive bands to detect DSFGs will be 225 and 280 GHz. If we consider the baseline noise from Table~\ref{tab:sourcesens}, since high-frequency channels are prone to high atmospheric loading, the $5\sigma$ detection cut is around 7.0 and 12.5 mJy for 225 and 280 GHz, respectively. Considering a sky fraction of 0.52 and the \citealt{Cai:2013} dusty models, we expect to detect around 32,000 and 36,000 DSFGs in each respective band. Both frequencies will yield a similar number of detections at the baseline noise, and both flux measurements help to constrain the spectral energy distribution of these galaxies. Combining these data with far-infrared measurements enables constraints on the physical properties of DSFGs, including redshift, temperature, emissivity, and luminosity  \citep{Su:2017, Reuter:2020, Bendo:2023}. SO will detect DSFGs up to redshifts $z=4$, and strongly lensed ones beyond $z = 6$, some with magnifications of several tens. The latter make promising follow-up targets for the Atacama Large Millimeter Array (ALMA), to resolve compact, young galaxies in their early evolutionary phases~\citep{spilker/etal:2016}.

\begin{figure}
    \centering    \includegraphics[width=0.49\textwidth]{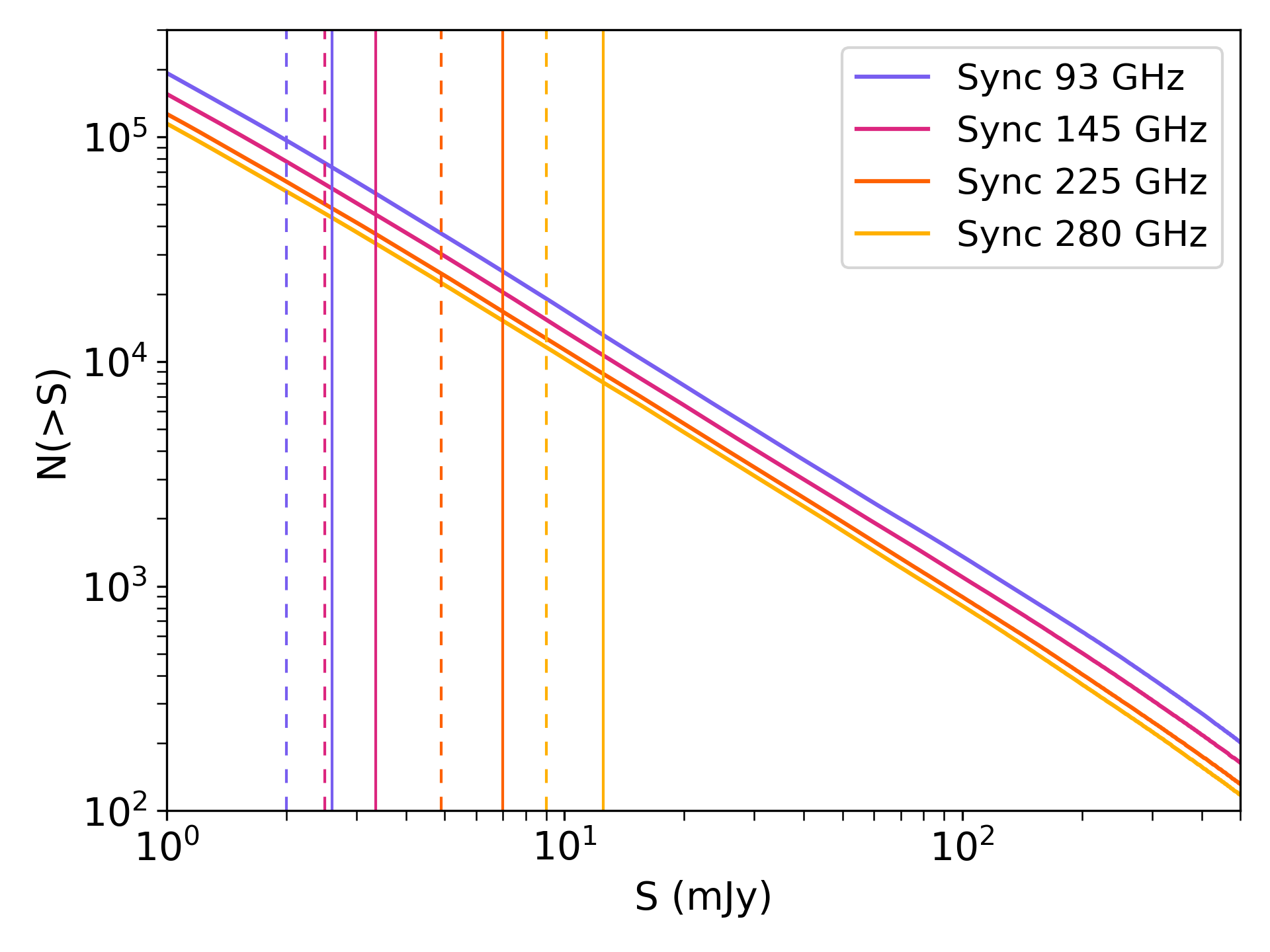}
\includegraphics[width=0.49\textwidth]{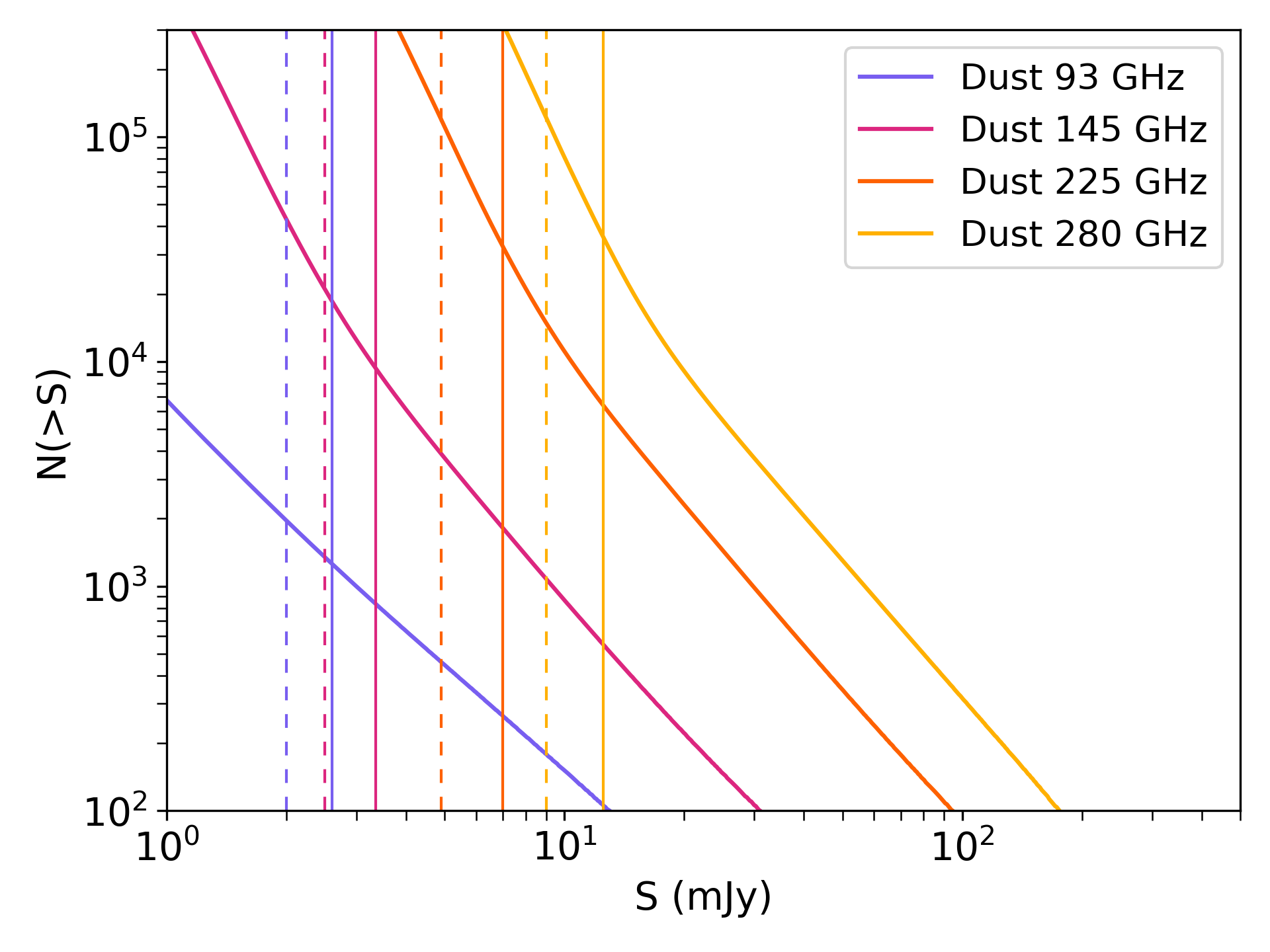}
    \caption{Total number of sources above different flux limits $S$, as a function of $S$ in mJy, assuming a sky fraction of 0.52. Left: models for AGN synchrotron emission at four frequencies \citep{Lagache:2020}.  Right: model dust emission for DSFGs \citep{Cai:2013} at the same four frequencies. Vertical lines indicate the $5\sigma$ sensitivity for the SO baseline (solid) and goal (dashed) sensitivities.}
    \label{fig:ngts}
\end{figure}

In Fig.~\ref{fig:ngts} we show the estimated number of AGN and DSFGs that we expect to detect given baseline or goal sensitivities. We note that the predicted number of DSFG detections varies by a factor of $\sim4$ between baseline and goal sensitivities. These forecasts are based on the best available models for the number of sources at different flux limits, but the low-flux regime is currently poorly constrained. SO data will refine these models and improve our understanding of AGN and DSFG populations as a whole. 

For the source sensitivities in Table \ref{tab:sourcesens}, backgrounds constitute a significant contribution to the total noise in all bands. At long wavelengths the relevant background is the CMB, but at shorter wavelengths the CIB is dominant. The possibility thus arises that deeper observations, reaching sensitivities below the CIB confusion limit, could be undertaken in sub-regions of the LAT survey, which would allow techniques such as probability of deflection approaches \citep[P(D); e.g.,][]{Glenn:2010} to set constraints on the CIB population beyond the confusion limit.

Most of the bright AGN found at millimeter frequencies are blazars. Blazars are AGN with relativistic jets aligned with the line of sight, often appearing variable. SO will increase the {\sl Planck} catalog of AGN \citep{PCCS2} by a factor of $\sim50$, and will provide regularly sampled light curves. For instance, the goal daily sensitivity at 93 GHz of 6.9\,mJy (Table \ref{tab:sourcesens}) translates to $3\sigma$ daily measurements of 21~mJy sources. According to the AGN model of \cite{Lagache:2020}, this implies that SO can provide light curves for 7,500 AGN at its regular observing cadence, expected to be every day or two when a source is in the field. (Each source will be unobservable for about a month out of each year, when it is too close to the Sun.) Combining multiple observations will increase the available number of light curves. For instance, averaging six observations would enable light curves of $\sim20,000$ AGN at the mean $3\sigma$ level.

This monitoring program will be a great improvement over the state of the art \citep{Richards:2011, Lister:2018, Bonato:2019, cortes_alma_2024}, and will enable study of the innermost portions of the jets, which are opaque at longer wavelengths.
Aside from measuring flaring events and stochastic variability, these data can be used to search for supermassive black hole binaries (SMBHBs) which are expected to have sinusoidal light curves. \cite{Kiehlmann:2024} estimate that about 1\% of the radio blazar population are SMBHBs. Hence, we expect to find $\mathcal{O}(100)$ SMBHBs with the monitoring program.

AGN light curves will be also valuable for cross-correlations with other tracers, including neutrino observatory measurements \citep[e.g.,][]{IceCubeCollaboration:2018}.
The blazar SED has two bumps. The low-energy bump \citep{Urry:1998}, spanning the radio to UV, peaks in the millimeter and is due to synchrotron emission from ultra-relativistic electrons.
The emission mechanism behind the high-energy bump, which spans the UV to gamma-ray portion of the electromagnetic spectrum, is still under debate. Its origin is either leptonic, inverse-Compton scattering of lower energy photons \citep{Maraschi:1992, Dermer:1993}, or hadronic, proton-synchrotron radiation or photo-pion production \citep{Mucke:2001, Dermer:2001}.
Cross-correlating blazar light curves over a range of wavelengths can elucidate the physical processes that shape and change blazar SEDs. 
Many studies have compared radio, optical, and gamma-ray blazar light curves \citep[e.g.,][]{Rani:2013, Liodakis:2018}.
However, few of these studies have included millimeter-wavelength light curves \citep[e.g.,][]{Hood:2023}, even though millimeter-wavelength emission is a strong indicator of synchrotron radiation.

\subsection{Insights into the polarized Galactic interstellar medium}
\label{sec:Galaxy}
\vspace{-0.05cm}

Arcminute-resolution millimeter-wave maps of the SO footprint, which includes much of the Galactic plane, will be valuable for answering many questions about the physical mechanisms responsible for polarized Galactic emission, via measurements of the dust and synchrotron spectral energy distributions across the sky, and detections or upper limits on polarized CO line emission and anomalous microwave emission \citep[AME;][]{SO_2022_Galactic_Science}.  
SO will detect AME polarization if the AME is polarized at the $\gtrsim$ 0.1\% level, which is sufficient to distinguish between competing theories of ultrasmall grain alignment \citep{SO_2022_Galactic_Science}, such as resonance paramagnetic alignment \citep{Lazarian:2000, Hoang:2013}, or suppression of alignment due to quantum mechanical effects \citep{Draine:2016}. Additionally, SO will determine what fraction of the AME could be contributed by magnetic dipole emission from ferromagnetic grains, which are distinguished by their unique polarization signature \citep{Draine:2013}. See \citet{Dickinson:2018} for a recent review. 
Together with large-area starlight polarization surveys such as PASIPHAE~\citep{tassis/etal:2018}, these data will yield new insights into the physics of Galactic emission and the multi-scale structure of the magnetized interstellar medium~\citep{hensley/etal:2019,fissel/etal:2019}. 

\begin{figure*}[t]
\center \includegraphics[width=\linewidth,trim={0cm 1cm 0 1cm},clip]{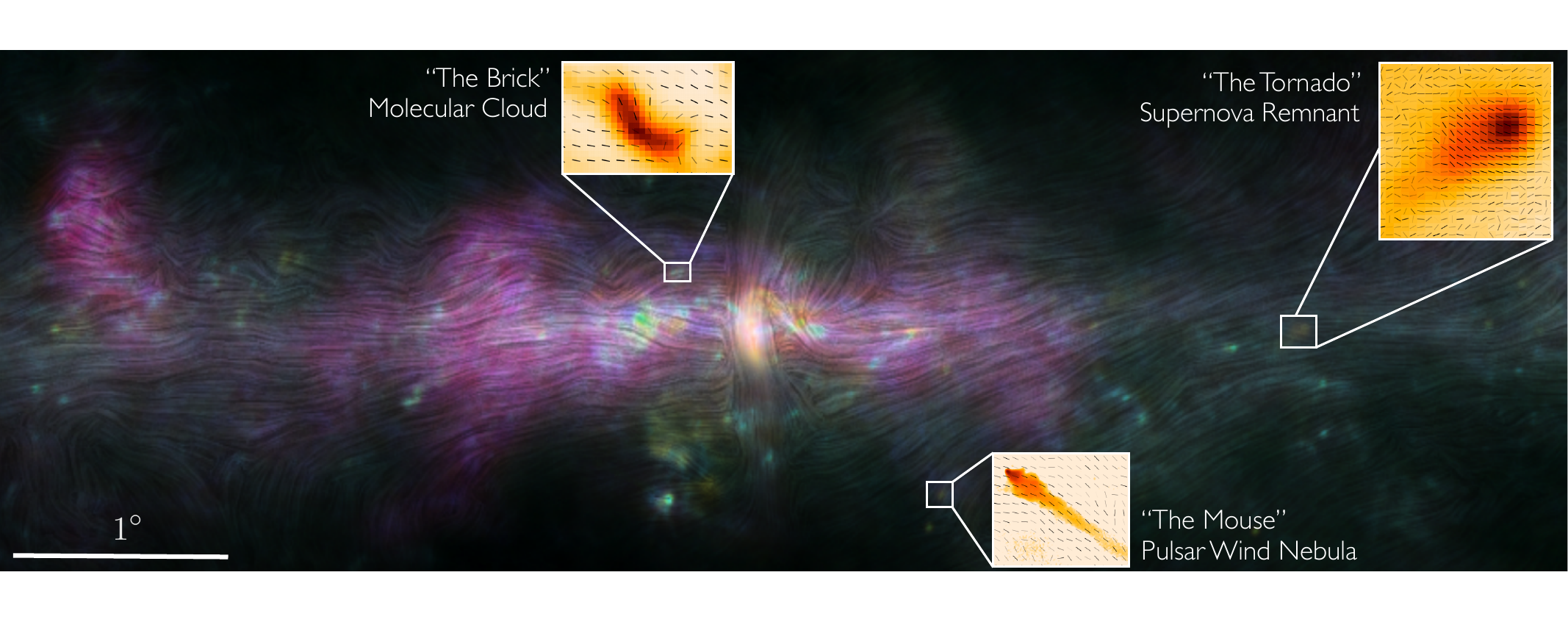}
\caption{ The ACT view of the Galactic Center and zoom-ins to particular regions \citep[adapted from][]{guan21}. The three-color background image comprises ACT 90 GHz (red), 150 GHz (green), and 220 GHz (blue) observations coadded with \PLANCK\ data at similar frequencies. Overlaid magnetic field orientation is from ACT 90 GHz polarization. The SO frequency coverage spans synchrotron-dominated (more red) to dust-dominated (more blue) emission, enabling a uniquely comprehensive view of the magnetic field morphology. Insets illustrate particular regions of interest. ``The Brick" Molecular cloud: {\it Herschel} total intensity overlaid with ACT 220 GHz magnetic field orientation data. ``The Tornado" supernova remnant: ACT 90 GHz intensity and magnetic field orientation data. ``The Mouse," a pulsar wind nebula: MeerKAT 1.28 GHz total intensity~\citep{Heywood:2019} and ACT 90 GHz magnetic field orientation. }
\label{fig:galaxy} 
\end{figure*}

Recent ACT observations of the Galactic center \citep[Fig.~\ref{fig:galaxy};][]{guan21} offer a glimpse of the physical processes probed at SO frequencies, which will complement multiwavelength observations \citep[e.g.,][]{Butterfield:2024, Heywood:2019}.  
As detailed in \citet{SO_2022_Galactic_Science}, SO will resolve the magnetic field structure of more than $860$ molecular clouds with 1 pc resolution and at least 50 independent, high-SNR polarization measurements per cloud, compared to {\sl Planck}'s sample of tens of clouds. Additionally, SO will be able to detect polarized dust emission in 400 Galactic cold clumps, prime candidates in the formation and evolution of stellar cores. This is a factor of 200 greater than \textit{Planck}, and will enable detailed statistical studies on their magnetic field structure and role in stellar formation \citep{Clancy:2023}. In the more diffuse ISM, the SO dust polarization maps can be used in conjunction with ISM emission tracers to quantitatively study the small-scale structure of polarized emission \citep[e.g.,][]{Clark:2015, Clark:2019, CordovaRosado:2024} and to probe the magnetic field structure in diffuse media \citep[][]{Lei:2024}.  

Component separation techniques and detector passband variation can be exploited to build maps of velocity-integrated CO line emission from continuum data \citep{planck2013:XIII}. \citet{SO_2022_Galactic_Science} forecast that SO will detect or constrain polarization of the CO(2--1) line at sub-percent levels in dense molecular clouds, including Ophiuchus and Orion. CO line polarization via the Goldreich-Kylafis effect probes the magnetic field strength and orientation in molecular clouds \citep{Goldreich:1981,Crutcher:2012}. 

\subsection{The composition of interstellar dust}

Classical models of the composition of interstellar dust posit two classes of dust: predominantly silicate and predominantly carbonaceous. These models predict a strong divergence of the dust total intensity and polarization spectra -- i.e., a wavelength-dependent polarization fraction -- in the frequency range probed by SO. 
However, analyses of data from \textit{Planck} and the Balloon-borne Large Aperture Submillimeter Telescope for Polarimetry (BLASTPol) find that the difference in spectral indices between dust total intensity and polarization at $\sim$submillimeter wavelengths is only on the order of a few percent at most \citep{Ashton:2018, Shariff:2019, Planck2018:XI}. 
This lack of evidence for the predicted divergence is driving new theoretical work on dust composition \citep{Guillet2018, Draine2021, Hensley:2023, Ysard:2024}. 
SO will test these  models, including constraining $\beta_\mathrm{dust}$, the spectral index of the modified blackbody SED characteristic of polarized dust emission. Using sky models from the Python Sky Model \citep{Thorne:2017}, we forecast that SO measurements combined with \PLANCK\ and {\sl Wilkinson Microwave Anisotropy Probe} ({\sl WMAP}) data will measure $\beta_\mathrm{dust}$ to $\sigma (\beta_\mathrm{dust}) = 0.005$, sufficient to test the two-component dust model~\citep{hensley/etal:2019}. Constraints on $\beta_\mathrm{dust}$ within smaller regions enable tests of recent evidence for variability of the dust spectral index \citep{SPIDER:2025} and of the dust SED more generally \citep{Pelgrims:2021, Ritacco:2023}. The evidence for $\beta_\mathrm{dust}$ variability can also be investigated in cross-correlation with filamentary ISM structures in neutral hydrogen \citep{BKHI}. 

Testing this paradigm will constrain the degree of dust homogenization via interstellar medium (ISM) processing, the rate of dust production in stars, and the composition of the grains that coagulate into solid bodies in protoplanetary disks \citep{hensley/etal:2019, Hensley:2023, Dartois:2024}. SO data can also be compared with optical starlight polarization measurements, including from PASIPHAE \citep{tassis/etal:2018}. The ratio of far infrared to optical polarization constrains the shape, porosity, and composition of interstellar grains~\citep{Planck2018:XII, Guillet2018, Draine:2021}.

\subsection{Solar system bodies and exo-Oort clouds}
\label{sec:planets}

Reflected sunlight from objects falls off rapidly with distance from the Sun, so the strongest limits on massive outer Solar System objects come from their intrinsic thermal emission.  The strongest bounds on unknown giant planets are set by the {\sl Wide-field Infrared Survey Explorer} ({\sl WISE}), excluding planets more massive than Saturn closer than 28,000 AU~\citep{luhman:2014}. The spectra of smaller, faster-cooling planets peak at longer wavelengths, well suited for millimeter observations (see~\cite{naess21a} for details on the model assumed here).  A $5~M_\oplus$ Planet 9 can be detected with SO, at distances ranging from 500 AU in the more shallowly observed regions over half the sky, to 900 AU over large parts of the sky. For a $5~M_\oplus$ Planet~9, a search with ACT data ruled out 17\% of the orbital parameter space ~\citep{naess21a}, and planetary ephemerides data have been used to exclude such an object within 500~AU~\citep{Fienga2020}.

The Solar System is thought to be surrounded by a large, roughly spherical shell of rock and dust known as the Oort cloud \citep{Oort:1950}. Our own Oort cloud has never been directly observed, but SO may be able to detect similar structures around other stars, i.e., exo-Oort clouds. \citet{SO_2022_Galactic_Science} forecast the pre-upgraded-LATR SO sensitivity to exo-Oort clouds at 280 GHz using simulated dust emission and Oort cloud emission profiles \citep{Baxter:2018} placed around \textit{Gaia} stars within 70\,pc \citep[see also][]{Nibauer:2020, Nibauer:2021}. SO will detect (at $\sim3\sigma)$ exo-Oort clouds if their occurrence rate is similar to the detection rate of giant planets \citep[][$f_{\rm Oort}$ = 0.3]{Fernandes:2019, Wittenmyer:2020}. As elaborated in \citet{SO_2022_Galactic_Science}, SO will also probe debris disks from exoplanet formation around nearby stars, constraining the statistical properties of the debris disk population and prioritizing candidates for higher-resolution follow-up with ALMA \citep[e.g.,][]{MacGregor:2017, Nederlander:2021}.

\subsection{Asteroid regoliths}

The composition and dynamics of asteroids provide an important window into the composition of the early solar system \citep[e.g.,][]{Michel2015}. Observations of asteroids at millimeter wavelengths have shown a consistent deficit in flux as compared to expectations from the infrared \citep[e.g.,][]{Johnston1982, Webster1988}. Since the unconsolidated surface, or regolith, of the asteroid is partially transparent in the millimeter, this millimeter flux deficit is connected to the composition of the regolith, which is otherwise difficult to ascertain. Historically this drop in flux has been interpreted as a change in effective emissivity due to regolith grain size or packing \citep{Redman1992}. However, more recent, comprehensive analysis has suggested the origin may lie instead in steep temperature gradients of several $\text{K}/\text{centimeter}$ lying just below the surface \citep{Keihm2013}.

Until recently, astronomers have been limited to targeted millimeter observations of asteroids. With the new generation of CMB experiments, however, it has become possible to make catalogs of asteroid flux in the millimeter. SPT made the first detection of three asteroids using a ground-based CMB telescope \citep{Chichura2022}, while ACT published the first extensive catalog of $170$ asteroids \citep{Orlowski-Scherer+23}. Both experiments found a systematic flux deficit in the millimeter, with \cite{Orlowski-Scherer+23} finding a spectral shape to the deficit wherein the deficit was more severe at $150$ and $220$\,GHz than at $90$\,GHz.

While the precise distribution of asteroid fluxes in the millimeter/sub-millimeter is currently unknown, to first order it will likely follow the size distribution, which is well described by a power law with $\alpha = -2.5$. Accounting for the $r^{2}$ scaling of the flux with asteroid radius, this leads to a $dN \propto F^{-1.25}$ scaling for $dN$ the number density of asteroids and $F$ the flux. Assuming this, we conservatively estimate that SO will detect approximately 100 times more than the \cite{Orlowski-Scherer+23} sample, on the order of ten thousand asteroids. We further expect the detection of the largest trans-Neptunian objects (TNOs). We scale the total S/N of a number of asteroids by
\begin{equation*}
\text{x}_{\text{Earth}}^{-2} \text{x}_{\text{Sun}}^{-1/2} \text{x}_{\text{diameter}}^2 \text{x}_{\text{S/N}}
\end{equation*}
where $\text{x}_{\text{earth}}$ is the ratio of the Earth-centered TNO distance to Earth-centered asteroid distance, $\text{x}_{\text{Sun}}$ is the same for the Sun-centered distance, $\text{x}_{\text{diameter}}$ is the ratio of their physical diameters, and $\text{x}_{\text{S/N}}$ is the projected SO/ACT S/N ratio. The exponents account for the scaling of the flux with the various parameters in the Rayleigh-Jeans limit of the Standard Thermal Model \citep{Lebofsky1986}.  Distances for the TNOs are evaluated January 1, 2025, and will not meaningfully change over the SO lifetime. What is not accounted for here is the emissivity of the TNO; performing this scaling for several asteroids helps marginalize over this uncertainty. We should detect a number of the largest TNOs, specifically (50000) Quaoar, (136472) Makemake, (136108) Haumea and (134340) Pluto at $\geq 5\sigma$ each. Other, larger TNOs have projected S/N $\simeq 1-3\sigma$, indicating that stacking analyses may also be fruitful.

\subsection{The unexplored millimeter transient sky} 
\label{subsec:transients}

The study of transient astronomical signals at millimeter wavelengths is opening a new frontier in astrophysics. Typically, the millimeter emission is synchrotron radiation produced by particles accelerated as a shock plows into an ambient medium. For decades, observations at centimeter wavelengths have been used to study the explosion and environment properties of explosive phenomena such as supernovae (SNe; e.g.,~\citealt{Weiler2002}), gamma-ray bursts (GRBs; e.g., \citealt{Chandra2012}), tidal disruption events (TDEs; e.g., \citealt{Alexander:2020}), and novae \citep[e.g.,][]{Chomiuk2021}. By contrast, the millimeter regime remains largely unexplored. 

Recent years have seen new opportunities for millimeter time-domain astronomy. One reason is that the discovery of young ($\lesssim$ days-old) transients by optical surveys has become routine, enabling fast turnaround follow-up observations using sensitive millimeter interferometers \citep{Ho:2019,Maeda:2021,Andreoni2022,Berger2023}. In addition, transients have been discovered blindly at millimeter wavelengths for the first time, in CMB survey data \citep{whitehorn/etal:2016,Guns:2021,naess21a, Li:2023, Biermann:2024}, and these data have also been used to put millimeter-wavelength flux limits on known extragalactic transients \citep{Hervias-Caimapo:2024}. 

With SO, transient events will be routinely and efficiently discovered directly in the millimeter, allowing for follow-up observations by the transient community. 
The SO rapid transient analysis pipeline will generate daily maps for transient detection and discovery, as well as light curves for each identified transient source to be released to the community within 30 hours. We will release light curves with a typical cadence of about one day, and a time resolution of order minutes. In addition, more slowly evolving faint transients can be detected in stacked maps. 

One class of extragalactic transients of interest are interacting supernovae, i.e., core-collapse supernovae exploding in dense circumstellar material. Growing evidence suggests that some massive stars shed a significant fraction of their mass in the final stages of their lives \citep{Smith:2014}, perhaps due to vigorous convection in the late stages of nuclear burning \citep{Quataert:2012, Shiode:2014} or binary interaction \citep[e.g., common-envelope evolution;][]{Chevalier:2012}. Enhanced mass loss on timescales of days to years would leave dense material close to the star (6--600\,AU). When the star explodes, this circumstellar material can be detected by its luminous millimeter-wave emission~\citep{Maeda:2021,Yurk2022,Maeda:2023}.

Another important class of extragalactic transients is long-duration GRBs (LGRBs), produced in the powerful jets launched by collapsing massive stars \citep{Piran2004}. 
A longstanding question in the GRB field is whether long-duration GRBs are a rare and distinct endpoint of stellar evolution, or the extremum of a broad continuum of relativistic stellar explosions (e.g., \citealt{Cano:2017}). SO could address this question by: (i) measuring the beaming fraction through the detection of off-axis events; (ii) finding events that bridge the gap between GRBs and ordinary supernovae, such as low-luminosity GRBs \citep[LLGRBs;][]{Cobb:2006,Liang:2007,Bromberg:2011,Nakar:2015}, which are a potential site of high-energy neutrinos \citep{Murase:2008} and cosmic rays \citep{Zhang:2018}; and (iii) looking for ``dirty fireballs'' \citep{Dermer:1999}, i.e., jets with lower initial Lorentz factors than classical GRBs. Inferring the initial Lorentz factor of a jet requires disentangling the forward and reverse shock components of the afterglow, which is most readily done using millimeter observations \citep{Perley:2014, Laskar:2018,laskar/etal:2019}.

GRB reverse shocks are predicted to be the most frequent class of extragalactic transients detected by SO (\citealt{Eftekhari2022}; Table \ref{tab:rates}). GRB reverse shocks are an emission component from shock propagation through the ejecta as the GRB jet collides with the interstellar medium. The reverse shock thus probes the physical properties of the GRB outflow \citep{Nakar:2004}. SO has the potential to conduct the most sensitive, unbiased survey of millimeter-wave reverse-shock emission over half the sky.

Relativistic jets launched in tidal disruption events, when a star is pulled apart by a supermassive black hole, can be luminous millimeter transients early in the event \citep{Zauderer:2011,Andreoni2022}. The fraction of TDEs harboring relativistic outflows appears to be small \citep{Alexander:2020}. SO blind millimeter-wave surveys will help to independently measure the event rate. In addition, millimeter observations are sensitive to outflows with lower energies than centimeter-wavelength observations \citep{Alexander:2020}.

SO transient observations could also probe the physics of shock acceleration. In the framework typically applied to model supernova centimeter-wave observations, the electrons are presumed to be accelerated into a power-law energy distribution (e.g., \citealt{Chevalier:1998}). Interestingly, millimeter observations of fast blue optical transients (FBOTs) such as AT2018cow \citep{Prentice2018} and AT2020xnd \citep{Perley:2021} show evidence for a separate emission component at millimeter wavelengths (100--200\,GHz) that does not fit the standard model used to describe the late-time low-frequency ($<40\,$GHz) data \citep{Margutti:2019,Ho:2019,Ho2022}. 
One possibility for the origin of the millimeter component is a Maxwellian distribution of electrons, i.e., electrons that are not accelerated into a power-law distribution \citep{Margalit2021,Margalit2024}.

\begin{table*}[t]
    \centering  
    \footnotesize
    \begin{tabular}{c|c|c|c|c}
    \hline\hline
      Class & Volumetric Rate & $L_\nu$ & Distance & Detection Rate\\
       & (yr$^{-1}$Mpc$^{-3}$) & (erg\,s$^{-1}$\,Hz$^{-1}$) & (Mpc) & (yr$^{-1}$)\\
      \hline
      Extragalactic fast (1--10\,d duration) & -- & -- & -- & $<10$ \\
      Long-duration GRB, on-axis & $4\times10^{-10}$ & $10^{32}$ & 1300 & 2 \\
      Long-duration GRB, off-axis & $6\times10^{-9}$ & $3\times10^{30}$ & 360 & 0.6 \\
      TDE, on-axis & $3\times10^{-11}$ & $10^{31}$ & 670 & 0.02 \\
    Low-luminosity GRB & $2\times10^{-7}$ & $10^{29}$--$10^{30}$ & 70--210 & 0.1--4 \\
    AT2018cow-like FBOT & $10^{-7}$ & $10^{30}$ & 210 & 2 \\
    Interacting SN & $10^{-8}$ & $10^{27}$--$10^{30}$ & $<210$ & $\lesssim0.2$\\
    \hline
    Stellar flares & -- & -- & -- & 180 \\
    \hline
    \end{tabular}
     \caption{Estimated detection rates by SO for different classes of extragalactic transients at 145 GHz, and for stellar flares at 90 GHz. We define off-axis long-duration GRBs to have $\theta_\mathrm{obs}=0.4$. 
     For transients lasting a few days (extragalactic fast and long-duration GRB on-axis) we use a goal sensitivity of 13~mJy rms in a single observation and require a 6$\sigma$ detection.
     Other events have a longer duration, so we use a baseline sensitivity of 4.9~mJy from one-week stacks and again require a 6$\sigma$ detection. Stacking maps would increase the sensitivity for longer-lived events. SO observations will constrain the significant theoretical uncertainty on the intrinsic rates of these transients. The detection rate of fast extragalactic transients is taken from systematic searches in SPT data (Guns et al. in prep). 
     The detection rate of stellar flares is extrapolated from recent ACT and SPT detections rate~\citep{Li:2023, Biermann:2024, Tandoi:2024} to the SO goal sensitivity and sky coverage, requiring $5\sigma$ detections. 
     }
    \label{tab:rates}
\end{table*}

SO will constrain the intrinsic rates of extragalactic millimeter transients, which are now quite theoretically uncertain for many classes of transients \citep{Eftekhari2022}. 
We forecast the distance that SO can probe for several key classes of extragalactic transients in Table~\ref{tab:goals}, with further forecasts in Table~\ref{tab:rates}. For most classes, this calculation simply depends on the luminosity of the source and the sensitivity of our observations. We estimate a baseline single-observation rms sensitivity of $\sim$13\,mJy at 145 GHz, and use this in our forecasts, assuming a $6\sigma$ threshold for extragalactic transient detection. We further forecast detection rates for each class of transients from current volumetric rate estimates and the SO detection distance, assuming that SO will discover transients over 52\% of the sky (the SO LAT survey footprint with the \textit{Planck} 80\% sky fraction Galactic plane mask applied). 

Projected detection rates are presented in Table \ref{tab:rates}. For fast events (durations between 1 and 10\,days), which would include some LGRB reverse shocks and on-axis LGRB afterglows, we use the upper limit on the rate computed by SPT (Guns et al. in prep). For the on-axis LGRB rate we use \citet{Lien2014}. For the LGRB and TDE luminosities at 145\,GHz we use theoretical predictions~\citep{Metzger:2015,Eftekhari2022} and caution that these are uncertain given the small number of observed events. We take the LLGRB rate from \citet{Soderberg2006}, and for the luminosity we use the observed values of GRB\,980425/SN\,1998bw \citep{Kulkarni:1998} and GRB\,171205A/SN\,2017iuk \citep{Perley:2017}. For AT2018cow-like FBOTs we use observed 100\,GHz luminosities \citep{Ho2022,Ho2023_AT2022tsd} and the volumetric rate from \citet{Ho2023}. For ordinary SNe we use an observationally constrained rate \citep{Li2011,Perley:2020}, and luminosities of the handful of events detected at millimeter wavelengths to date \citep{Berger2023}. We find that ordinary SNe would likely only be detected out to a few Mpc, and thus we do not list these in Table \ref{tab:rates}, as we expect the detection rate for extragalactic normal SNe to be low. For interacting SNe we use a volumetric rate from the Zwicky Transient Facility \citep{Perley:2020} and a range of luminosities from the modeling in \citet{Yurk2022}.

SO will also blindly discover and characterize many stellar flares. Recently, the ACT \citep{naess21b, Li:2023, Biermann:2024} and SPT \citep{Guns:2021, Tandoi:2024} collaborations published blind discoveries of bright transients. We use these to extrapolate the expected stellar flare detection rate for SO with the fully populated LATR, assuming the goal single-observation sensitivity at 93 GHz from Table \ref{tab:sourcesens}. Assuming that the number density of events is proportional to the source flux density $S^{-3/2}$ and extrapolating from \citet{Tandoi:2024}, we estimate that SO will detect at least $\sim120$ flaring events per year at $5\sigma$ at Galactic latitudes $|b| > 5^\circ$. The scale height and flaring rate of stars varies with stellar type \citep{Aganze:2022, Murray:2022}, but the intrinsic rate of these events is higher closer to the Galactic plane. However, the expected stellar flare detection rate close to the Galactic plane is difficult to extrapolate from ACT and SPT results, as both of those surveys had much shallower scan depth near the plane. Estimating the detection rate at $|b| < 5^\circ$ by extrapolating from \citet{Li:2023}, we estimate that SO will detect at least 50\% more events at these low Galactic latitudes. We thus forecast a $5\sigma$ stellar flare detection rate for the fully populated SO LATR of at least $\sim 180$ per year.
 
With daily updates on source light curves with high time resolution over a wide field, SO will produce a large catalog of flaring stars for investigations of stellar physics. ACT also recently reported an observation of a classical nova \citep{Biermann:2024}, one of only a few observations of nova outbursts at millimeter wavelengths \citep{Ivison:1993,Nielbock2003,Chomiuk2014,Diaz2018}.

SO will complement a number of concurrent multiwavelength time-domain surveys. During SO operations, high-energy missions including the Space Variables Object Monitor \citep{Cordier:2015} and Einstein Probe \citep{Yuan:2015} will survey the sky in soft X-rays. For transients discovered by SO, these experiments could provide limits on the presence of a high-energy counterpart. In addition, ULTRASAT \citep{Shvartzvald2023} and UVEX \citep{Kulkarni2021} will conduct the first wide-field high-cadence surveys in the UV; ULTRASAT will spend a substantial fraction of its observing time in the southern hemisphere. Rubin will provide a wide-field, low-cadence, sensitive multi-band survey. Rubin observations of supernova mass-loss can be combined with SO observations of the terminal explosion. Several wide-field surveys will cover the entire sky to a depth comparable to 2MASS \citep{Moore:2019}, including WINTER \citep{Lourie:2020}. Several sensitive wide-field radio transient surveys are planned for the next decade, including the Deep Synoptic Array 2000 \citep{Hallinan:2019}, the ASKAP Variables and Slow Transients Survey~\citep{Murphy:2013}, and the Square Kilometer Array~\citep{Fender:2015}.

SO adds an important dimension to these complementary experiments with its sensitivity to polarization, which is a useful diagnostic for distinguishing between different classes of sources. The first extragalactic transient found in a blind millimeter survey was linearly polarized with a polarization angle that changed over the duration of the burst~\citep{whitehorn/etal:2016}. Such behavior is consistent with emission from a jet, and thus the source may have been a GRB afterglow. 

Finally, we note that the wide, blind nature of the SO time-domain survey leaves open the possibility of discovering a wholly new class of transient events.

%% file: tex/Summary.tex
In this paper, we have described planned infrastructure enhancements relevant to the wide-area survey that will be undertaken with the SO LAT.  The fully-populated SO LATR will include four additional MF and two additional UHF optics tubes, as described in Sec.~\ref{sec:infrastructure}.  These additions will nearly double the mapping speed over the existing configuration, with the full instrument expected to begin observations in 2028.  The final co-added map depth at the conclusion of the survey in 2034 will reach $2.6$~$\upmu$K$\cdot$arcmin, over roughly 61\% of the sky.  In addition to the new detectors, the planned infrastructure also includes a new photovoltaic power system, supplying 70\% of the power needs of the observatory, as well as an improved data pipeline that will facilitate map delivery to the community and detection of millimeter-wave astrophysical transients.  The properties of the fully-populated LATR are summarized in Table~\ref{tab:sens}.

The science goals and forecasts for the wide-area survey conducted with the fully-populated LATR are described in Sec.~\ref{sec:science} and summarized in Table~\ref{tab:goals}.  The forecasts presented in this work include only statistical errors; our understanding of systematic errors will be refined in future work, particularly as the instrument is now taking data.  These include improved constraints on the scale dependence and Gaussianity of the primordial perturbations (reaching $\sigma(f_{\rm NL}^{\rm loc}) = 1$); improved constraints on new light relativistic species (reaching $\sigma(N_{\rm eff}) = 0.045$); tight constraints on the sum of the neutrino masses ($\sigma(\sum m_\nu) = 0.03$~eV, or 0.015~eV in combination with {\sl LiteBIRD}); percent-level constraints on the amplitude of density fluctuations at redshifts $1 < z < 2$ via CMB lensing cross-correlations with LSS surveys; and a tSZ-selected galaxy cluster sample comprising 33,000 objects, including $\approx 200$ at $z > 2$.  In general, the high-SNR component-separated blackbody temperature and Compton-$y$ maps, as well as the reconstructed gravitational lensing maps, from the complete SO LAT survey will enable a broad range of cosmological and astrophysical science, which we have only briefly covered in this paper.  As evidenced by progress in cosmology in recent decades, a data set this rich will enable additional science that has yet to be envisioned or forecast here.

SO will also catalog $\sim100,000$ AGN and provide high-SNR light curves for $\mathcal{O}(10,000)$ blazars; map the magnetic field structure of hundreds of Galactic molecular clouds; constrain the composition of interstellar dust; detect or strongly constrain the presence of a Planet 9 (with the ability to rule out a 5~$M_{\oplus}$ Planet 9 out to 500-900 AU); detect or place limits on the population of exo-Oort clouds; and measure the thermal emission from thousands of asteroids, enabling new statistical investigation of asteroid regoliths. Finally, SO will carry out the largest blind survey of transient phenomena in the millimeter to date, which will detect and characterize GRBs, TDEs, FBOTs, and supernovae, as well as on order of a hundred stellar flares per year. The sensitive, large-area maps of millimeter-wavelength emission in both total intensity and linear polarization will enable a wealth of Galactic and extragalactic science, and will open new discovery space in the time domain.

\begin{acknowledgments}
This work was supported by the National Science Foundation (Award No.~2153201, UEI GM1XX56LEP58). This work was supported in part by a grant from the Simons Foundation (Award \#457687, B.K.).  This work was supported in part by a Laboratory Directed Research and Development award from SLAC National Accelerator Laboratory under Department of Energy Contract No. DE-AC02-76SF00515. The research was carried out in part at the Jet Propulsion Laboratory, California Institute of Technology, under a contract with the National Aeronautics and Space Administration (80NM0018D0004). This document was prepared by Simons Observatory using the resources of the Fermi National Accelerator Laboratory (Fermilab), a U.S. Department of Energy, Office of Science, Office of High Energy Physics HEP User Facility. Fermilab is managed by Fermi Forward Discovery Group, LLC, acting under Contract No.~89243024CSC000002.  F.~Nati acknowledges funding from the European Union (ERC, POLOCALC, 101096035). Views and opinions expressed are however those of the authors only and do not necessarily reflect those of the EU or the ERC. Neither the EU nor the granting authority can be held responsible for them. MH acknowledges financial support from the National Research Foundation of South Africa.  JCH acknowledges support from the Sloan Foundation and the Simons Foundation. SEC acknowledges support from the Sloan Foundation. ADH acknowledges support from the Sutton Family Chair in Science, Christianity and Cultures, from the Faculty of Arts and Science, University of Toronto, and from the Natural Sciences and Engineering Research Council of Canada (NSERC) [RGPIN-2023-05014, DGECR-2023-00180]. This work was supported by a grant from the Simons Foundation (CCA 918271, PBL). AC acknowledges support from the STFC (grant numbers ST/W000997/1 and ST/X006387/1). EH is supported by a Gates Cambridge Scholarship (Grant No. OPP1144 from the Bill \& Melinda Gates Foundation).  The SISSA group acknowledges partial support by the Italian Space Agency LiteBIRD Project (ASI Grants No. 2020-9-HH.0 and 2016-24-H.1-2018), as well as the InDark and LiteBIRD Initiative of the National Institute for Nuclear Physics, and the RadioForegroundsPlus Project HORIZON-CL4-2023-SPACE-01, GA 101135036. MM acknowledges support from NSF grants AST-2307727 and  AST-2153201 and NASA grant 21-ATP21-0145. This work was supported in part by United Kingdom Research and Innovation (UKRI) and the Science and Technology Facilities Council (STFC) (Grant no. ST/X006344/1).    CHC acknowledges ANID FONDECYT Postdoc Fellowship 3220255 and BASAL CATA  FB210003.  NS acknowledges support from DOE award number DE-SC0025309.

We thank Patricia Diego-Palazuelos for facilitating code to run the isotropic cosmic birefringence estimation.  Some of the results in this paper have been derived using the {\tt healpy}~\citep{Zonca2019} and {\tt HEALPix} package~\citep{Gorski2005}, as well as the {\tt PySM} package~\citep{Thorne:2017,Zonca_2021, ThePan-ExperimentGalacticScienceGroup:2025}.

\end{acknowledgments}

%% file: tex/Sensitivities.tex
Table~\ref{tab:sens} provides map-domain sensitivities for the full-depth maps expected at the conclusion of the SO LAT survey in 2034.  To obtain these numbers, we take a similar approach to that in~\cite{so_forecast:2019}, using an atmospheric noise model informed by ACT observations combined with detector NET sensitivities estimated using {\tt bolocalc}~\citep{hill/etal:2018}. For the nominal-survey SO forecasts described in~\cite{so_forecast:2019}, it was assumed that 20\% of the data collected over a 5-year duration would be usable for cosmological analysis (after accounting for uptime, CMB field availability, and data quality cuts, all of which were estimated based on historical data from ACT).  For this work, we use the same detector NET sensitivities and 20\% observation efficiency, but make the following changes.  Rather than assuming a 5-year survey with the OT configuration from \cite{so_forecast:2019} (comprising 1 LF, 4 MF, and 2 UHF OTs), we assume 3 years of nominal SO observations with that tube configuration, followed by 6 years of observations with the fully populated LATR (comprising 1 LF, 8 MF, and 4 UHF OTs, as described in Sec.~\ref{sec:infrastructure}).  \cite{so_forecast:2019} assumed that the sky fraction observed would comprise  $f_{\rm sky}=0.47$, with a post-masking footprint of $f_{\rm sky}=0.4$ available for cosmological analysis (avoiding regions of Galactic contamination and poor cross-linking near the footprint edges).  In this work, following updates to the SO scan strategy that accommodate a wider range of science goals, we assume that throughout the 9-year survey duration the SO LAT will map a sky fraction $f_{\rm sky}=0.61$.  These assumptions lead to the full-depth map sensitivities given in Table~\ref{tab:sens}.  For completeness, we provide the ``goal'' and ``baseline'' per-OT NET sensitivities in Table~\ref{tab:sens_NET}.  Further details for the cosmological science forecasts and the point-source and transient forecasts are given below.

\begin{table*}[t]
\begin{center}
    \caption{SO Large Aperture Telescope NET per Optics Tube}
    \begin{tabular}{c | c | c | c | c | c | c }
    \hline
        Band & LF1 & LF2 & MF1 & MF2 & UHF1 & UHF2 \\
        \hline
        Frequency [GHz] & 27 & 39 & 93 & 145 & 225 & 280 \\
        Baseline NET [$\upmu$K$\sqrt{\rm s}$] & 48 & 24 & 10.8 & 13.4 & 21.2 & 50.9 \\
        Goal NET [$\upmu$K$\sqrt{\rm s}$] & 35 & 18 & 7.8 & 8.4 & 14.1 & 35.4
    \end{tabular}
\begin{tablenotes}
\item Forecasts in this work and in~\cite{so_forecast:2019} are based on ``baseline'' and ``goal'' noise models.  This table provides the associated NET per OT; note that each OT includes three $\sim$150-mm detector arrays.  The baseline model represents the requirements for SO, while goal represents the expected performance.  These models include parameters to encapsulate the key drivers for instrumental sensitivity, including: the emission from and opacity of the atmosphere; the properties of the telescope (emission, spill, etc.); properties of the receiver and cryogenic optics (emission, losses, and reflections from the windows, lenses, and cryogenic cold stop); properties of the filters that define the passbands; and detector properties (yield, noise, etc.).  The sensitivities presented here were computed at the start of the design process, when many instrumental details were yet to be finalized.  For this reason, the parameter choices are uninformative and are not reported here.  Pre-deployment testing of the MF OTs indicates performance exceeding the baseline requirement and consistent with goal performance~\citep{Sierra:2025}.
\end{tablenotes}
\label{tab:sens_NET}
\end{center}
\end{table*}

{\it Cosmology:} All of our cosmological science forecasts use the ``goal'' sensitivity levels and assume that the sky fraction available for analysis after masking the Galaxy is $f_{\rm sky}=0.4$.  However, over this sky fraction, we have assumed slightly different map-domain sensitivities of $\{39, 20, 3.5, 3.8, 9.1, 22\}$~$\upmu\mathrm{K} \cdot \mathrm{arcmin}$ at $\{27, 39, 93, 145, 225, 280\}$~GHz, respectively, compared to the values of $\{44, 23, 3.8, 4.1, 10, 25\}$~$\upmu\mathrm{K} \cdot \mathrm{arcmin}$ given in Table~\ref{tab:sens}, due to different survey-strategy assumptions that were made earlier in the cosmological analyses.  These small differences in the map-domain noise levels are not expected to significantly change our cosmological forecasts at the precision level quoted in this work.

{\it Transient and point source science:} We use two types of sensitivities for our transient and point source forecasts. First, we estimate our sensitivities to point sources from a single observation, that is, for a point on the sky that drifts once through the entire focal plane, which is scanning in azimuth at a fixed elevation. We base these sensitivities on simulated maps made with our nominal observing strategy, in which the majority of observations occur at $40^{\circ}$ elevation (70.6\% of observations), with a minority at $50^{\circ}$ (15.7\%) and at $60^{\circ}$ (13.7\%). We use a fixed set of detector NETs for these simulations, and we then rescale the maps to correspond to the NETs used for the full-depth cosmological forecasts listed in Table~\ref{tab:sens}.  We take the median map depth, in units of $\upmu\mathrm{K} \cdot \mathrm{arcmin}$, for each of the observing elevations, and calculate their weighted mean based on the observing fraction at each elevation.  Second, we estimate our sensitivities to point sources in the final, full-survey maps by spreading the array sensitivity accumulated over the survey duration over the survey area, using $f_{\rm sky} = 0.61$ and an observing efficiency of $20\%$, as described above.  For both single-observation and full-depth maps, we obtain point source sensitivities by converting the map depth in $\upmu\mathrm{K} \cdot \mathrm{arcmin}$ to mJy using the beam FWHMs listed in Table~\ref{tab:sens} and applying an inverse-variance weighting with an atmospheric $1/\ell$ noise spectrum with a power-law index of $-3.5$ and $\ell_{\rm knee} = \{500, 500, 2100, 3000, 3800, 3800\}$ for $\{27, 39, 93, 145, 225, 280\}$~GHz, respectively. Here, the $\ell_{\rm knee}$ values are estimated based on ACT data.\footnote{Note that \citet{so_forecast:2019} also made use of a set of $\ell_{\mathrm{knee}}$ values for the noise model used in forecasts (which is used in the cosmological forecasting in this paper as well), but there the term would have been better called a \textit{reference} multipole, $\ell_{\mathrm{ref}}$, rather than a \textit{knee} multipole, which is usually defined as the multipole where the red noise caused by the atmosphere reaches the same level as the white noise; \citet{so_forecast:2019} instead defined a ``red noise'' amplitude, $N_{\mathrm{red}}$, to scale the reference multipole.  The relationship between the two is: $N_{\mathrm{red}} / \ell_{\mathrm{ref}}^{\alpha} = \sigma^2 / \ell_{\mathrm{knee}}^{\alpha}$, where $\alpha$ is the power-law index of the red noise. For the point source forecasts in this paper, our choices of $\ell_{\mathrm{knee}}$ and $\alpha$ are independent of the model used in \citet{so_forecast:2019}, but are still chosen to be consistent with the atmospheric noise measured by ACT.}  For the full-depth maps, we include additional ``noise'' contributions from the CMB and the CIB, since these terms dominate over the atmospheric noise in this case (other background sources like unresolved AGN are ignored). We ignore these terms for the single-observation maps since they are subdominant. The resulting sensitivities for both the single-observation and full-survey cases are given in Table~\ref{tab:sourcesens}.